# Experiment Neutrino-4 search for sterile neutrino and results of measurements

A.P. Serebrov[1*], R.M. Samoilov[1], V.G. Ivochkin[1], A.K. Fomin[1], V.G. Zinoviev[1], P.V. Neustroev[1], V.L. Golovtsov[1], S. S. Volkov[1], A.V. Chernyj[1], O.M. Zherebtsov[1], M.E. Chaikovskii[1], A.L. Petelin[2], A.L. Izhutov[2], A.A.Tuzov[2], S.A. Sazontov[2], M.O. Gromov[2], V.V. Afanasiev[2], M.E. Zaytsev[1, 3], A.A.Gerasimov[1], V.V. Fedorov[1]

[1]*National Research Center Kurchatov Institute – Petersburg Nuclear Physics Institute, 188300 Gatchina, Russia*
[2]*JSC "State Science Center Research Institute of Atomic Reactors", 433510 Dimitrovgrad, Russia*
[3]*Dimitrovgrad Engineering and Technological Institute MEPhI, 433511 Dimitrovgrad, Russia*

Abstract

The experiment Neutrino-4 had started in 2014 with a detector model and then was continued with a full-scale detector in 2016 - 2021. In this article we describe all steps of preparatory work on this experiment. We present all results of the Neutrino-4 experiment with increased statistical accuracy provided to date. The experimental setup is constructed to measure the flux and spectrum of the reactor antineutrinos as a function of distance to the center of the active zone of the SM-3 reactor (Dimitrovgrad, Russia) in the range of 6 - 12 meters. Using all the collected data, we performed a model-independent analysis to determine the oscillation parameters $\Delta m_{14}^2$ and $\sin^2 2\theta_{14}$. The method of coherent summation of measurement results allows to directly demonstrate the oscillation effect. We present the analysis of possible systematic errors and the MC model of the experiment, which considers the possibility of the effect manifestation at the present precision level. As a result of the analysis, we can conclude that at currently available statistical accuracy we observe the oscillations at the $2.9\sigma$ level with parameters $\Delta m_{14}^2 = (7.3 \pm 0.13_{st} \pm 1.16_{syst})\,\text{eV}^2 = (7.3 \pm 1.17)\,\text{eV}^2$ and $\sin^2 2\theta = 0.36 \pm 0.12_{stat}(2.9\sigma)$. Monte Carlo based statistical analysis gave estimation of confidence level at 2.7σ. We plan to improve the currently working experimental setup and create a completely new setup in order to increase the accuracy of the experiment by 3 times. We also provide a brief analysis of the general experimental situation in the search for sterile neutrinos.

---

[*]serebrov_ap@pnpi.nrcki.ru

## I. INTRODUCTION

Experiments on search for possible neutrino oscillations in sterile state have been carried out for many years. There are experiments at accelerators, reactors, and artificial neutrino sources [1-28]. A sterile neutrino can be considered as a candidate for the dark matter particles.

By combining results of various reactor experiments one can estimate the ratio of the observed antineutrino flux to the predicted flux to be 0.927±0.023 [29-32]. The deviation from no oscillation hypothesis is about 3 standard deviations. This level is not yet sufficient to have confidence in existence of the reactor antineutrino anomaly. Importantly, the method to test hypothesis of oscillation into sterile state, in which one compares the measured antineutrino flux with the expected flux from the reactor is not satisfactory, because of the problems with accurate estimation of both a reactor antineutrino flux and efficiency of an antineutrino detector. The possible process of oscillations to a sterile state at small distances from an active zone of the reactor is shown in FIG. 1, which was published in [32].

If oscillation process does exist, then deviation of antineutrino flux from flux calculated in assumption of no oscillation can be described by the equation:

$$P(\bar{v}_e \rightarrow \bar{v}_e) = 1 - \sin^2 2\theta_{14} \sin^2\left(1.27\frac{\Delta m_{14}^2[eV^2]L[m]}{E_{\bar{v}}[MeV]}\right), (1)$$

where $E_{\bar{v}}$ is antineutrino energy in MeV, L – distance in meters, $\Delta m_{14}^2$ is difference between squared masses of electron and sterile neutrinos, $\theta_{14}$ is mixing angle of electron and sterile neutrinos. The experimental test of the oscillation hypothesis requires measurements of the antineutrino flux and spectrum as near as possible to a practically point-like antineutrino source.

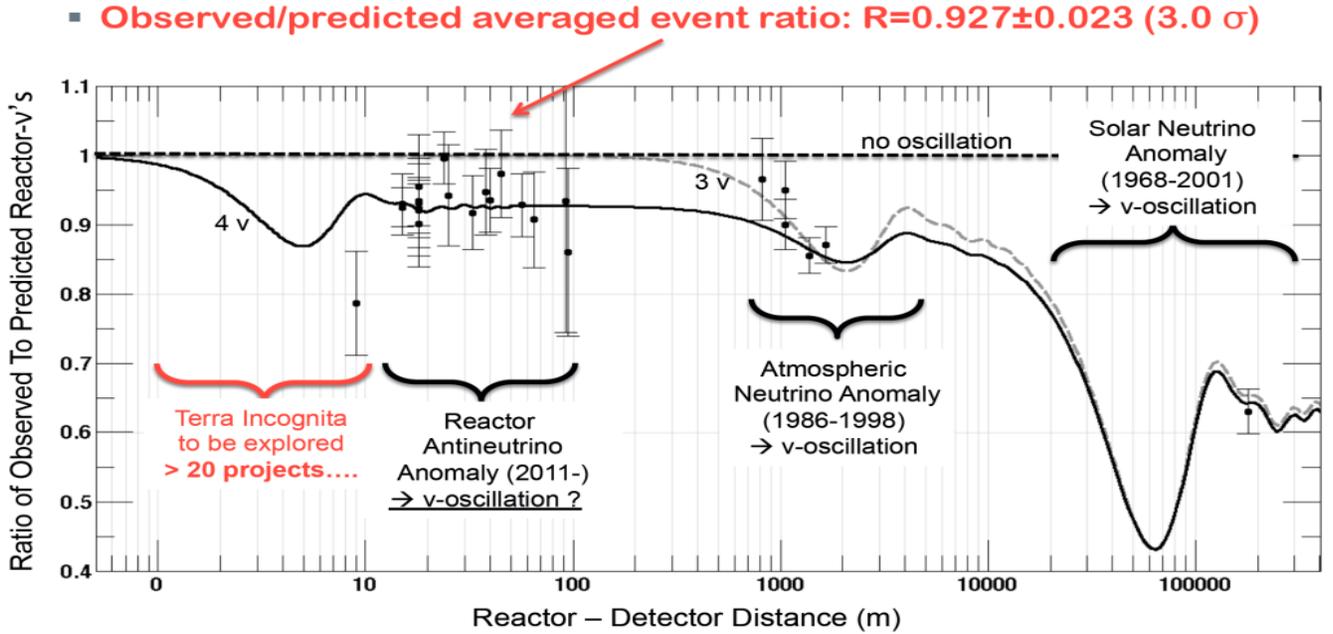

FIG. 1. The possible process of oscillations to a sterile state at small distances from an active zone of the reactor.

Based on equation (1), the oscillation hypothesis can be verified by direct measuring of distance dependences of antineutrino flux and spectrum at as short as possible distances to practically point-like antineutrino source. The oscillations manifest themselves in two effects: the way neutrino flux varies with distance deviate from the quadratic decrease form $1/L^2$; neutrino spectrum varies with distance. Therefore, a neutrino detector has to be movable and a spectrum sensitive. Our experiment focuses on the task of confirming possible existence of a sterile neutrino at a certain confidence level or disproving it. We have studied potential of research reactors in Russia to conduct new experiments. The research reactors should be employed for performing such experiments, since they possess a compact reactor core, so that a neutrino detector can be placed at a sufficiently small distance from it. Unfortunately, research reactor beam halls have quite a large background of neutrons and gamma quanta from the operating reactor, which makes it difficult to perform low background experiments. Due to some peculiar characteristics of its construction, reactor SM-3 provides the most favorable conditions for conducting an experiment on search for neutrino oscillations at small distances. At the same time, the SM-3 reactor, like other research reactors, is located on the Earth surface, hence an experimental setup of neutrino experiment is exposed to high cosmic background and it appears to be the major difficulty for the experiment.

Here is the structure of the article and brief description of the sections.

The first part of the article describes preparations to the experiment, which includes the study of the background at the reactor SM-3. The antineutrino signal should be selected from the background produced by cosmic rays, which is inevitably present. Antineutrino penetrates the biological shielding of the reactor without weakening. The method of the antineutrino signal selection is to compare the results obtained with operating and stopped reactor. This is the so-called ON-OFF count rate. That method can be applied if the background conditions in vicinity of the antineutrino detector do not change when the reactor is switched between operating and stopped states. The most dangerous source of the background is the fast neutrons which can simulate the antineutrino registration signal. This issue is addressed in sections II-V where we present measurements carried out with the detector of fast neutrons, background suppression, building of the passive shielding, and investigation of its characteristics.

Studies of the cosmic ray background were carried out with a test model of a neutrino detector filled with 400 liters a gadolinium doped liquid scintillator (0.1% Gd), and covered with the active shielding against cosmic muons. They are presented in sections VI, VII. In these studies was obtained the measurements of fluctuations of the cosmic rays background with time inside the neutrino channel (passive shielding), where the antineutrino detector moves. The model of the detector was applied to investigate the method of antineutrino registration as a result of the inverse beta decay (IBD) process, as well as the effectiveness of active shielding to suppress the background of cosmic rays.

The next stage of research (section VIII) was carried out on a sectioned model of the detector, also as a next preparatory stage for the creation of the full-scale detector. Here, an attempt has been made to separate the neutrino signal from the fast neutron signal, relying on the fact that in the inverse beta decay reaction there are two gamma quanta (511keV) that can be recorded in adjacent sections. Also we present the results of studies with a single section.

Sections IX-XVI are devoted to the full-scale detector: energy calibration of the detector, the computer model of the detector, obtaining the spectrum of antineutrino signals and comparing it to the calculated spectrum.

In sections XVII, XVIII and XIX we discuss the spectrum independent method for analysis of the neutrino signal and the Monte Carlo simulations of this method. This is an extremely important point that allows us to move to real measurements and to process the results in order to reveal possible neutrino oscillations at short distances.

Sections XX - XXV are devoted to the results of the measurements, their analysis, searching for the oscillation signal, statistical and systematic uncertainties of the oscillation effect.

In section XXVI we describe MC-modeling of the experiment which taking into consideration obtained statistical accuracy.

The final part of the article (Section XXVII) is devoted to the comparison of the results of this study with the results of other experiments. This analysis is of great importance because it insures that the obtained result does not contradict the results of other experiments on search for sterile neutrinos.

## II. REACTOR SM-3

Initially, the SM-3 reactor having maximum power 100 MW was designed for carrying out both beam and loop experiments. Five beam halls were built, separated from each other with 1 m wide concrete walls (FIG. 2).

This enabled carrying out experiments on neutron beams, without changing background conditions at neighboring installations. Later on, the main experimental program was focused on the tasks concerned with irradiation in the reactor core center. For 25 years of exploitation, a significantly high fluence was accumulated in materials of the reactor pressure vessel, which necessitated its replacement. Setting a new reactor pressure vessel on old reactor core barrel without joints with horizontal reactor beamlines was the simplest way to solve this problem. This decision led to raising of the reactor core center by 67 cm relative to previous position. As a result, horizontal beamlines ceased to be used, as priority was given to conducting loop experiments. Neutron flux in the location of the former beamlines was lowered by four orders of magnitude. Therefore, neutron background (thermal neutrons) decreased to level about $4 \cdot 10^{-3} \ n/(cm^2 s)$ in the former beam halls. It is several orders of magnitude lower than a typical neutron background in the beam hall of a research reactor.

Besides the favorable level of the reactor background, another advantage of the SM-3 reactor is its compact reactor core ($35 \times 42 \times 42 cm^3$), with high reactor power equal to 90 MW.

In one of the former beam halls we created a laboratory to carry out an experiment on search for oscillation of reactor antineutrino into a sterile state. This hall fulfills conditions important to our experiment: small distance (5 m) from the reactor core to the wall of the hall; size of the hall enables to carry out measurements of antineutrino flux in wide range 6 – 12 m.

**Reactor:** SM-3 reactor in Dimitrovgrad (Russia): 100 MW compact core 35x42x42 cm$^3$

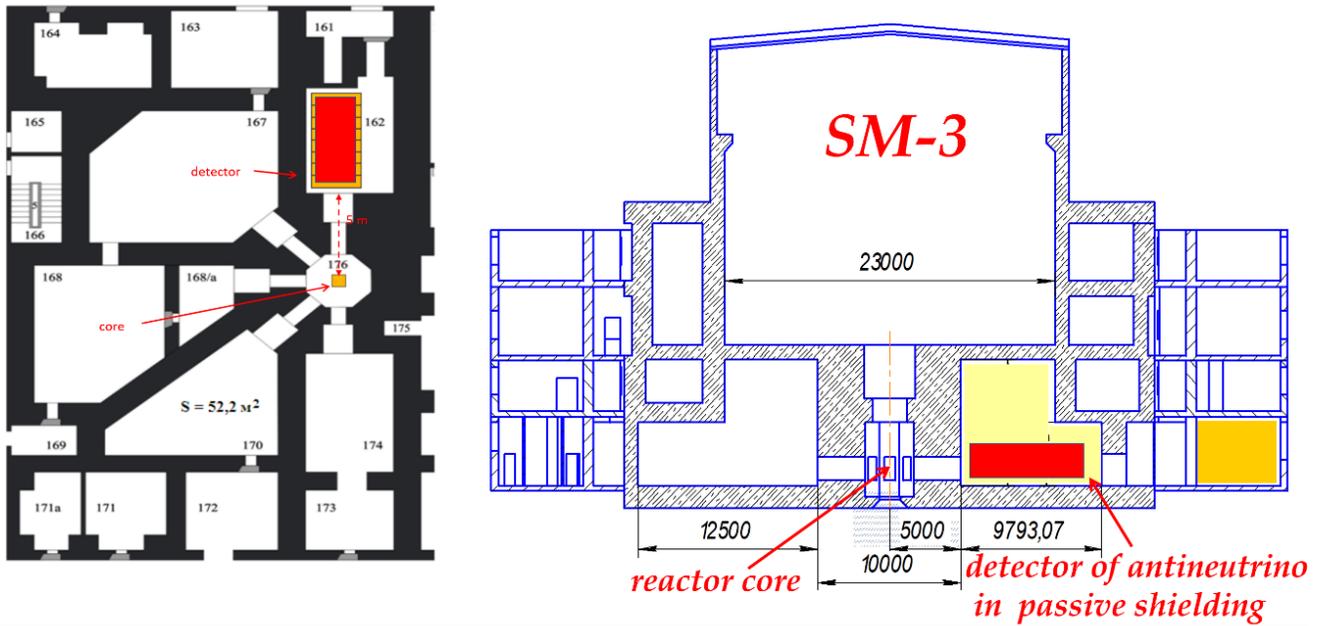

FIG. 2. Detector location at the SM-3 reactor.

In making preparations of the laboratory, the slide valve of the former neutron beamline has been upgraded to close all possible gaps to stop neutrons and gammas. As a result, the background of fast neutrons has decreased to the level of $10^{-3}$ $n/(cm^2 s)$, i.e. practically, to the level of neutron background on the Earth surface, caused by cosmic rays. Achieved conditions can be considered as the most favorable of all possible for experiment on search for neutrino oscillation at small distances.

Up to $10^3$ neutrino events are expected to occur per day, at the reactor with nominal power 90 MW at 8 m distance from the reactor core center, with detector volume of 1 m$^3$. However, registration efficiency in our method is only about 30%, so with 1 m$^3$ of liquid scintillator, we can record about 300 antineutrino events per day. This event rate is considered to be not very high, but it is sufficient to carry out experiment with cosmic background conditions. The scheme of antineutrino detector placement at the SM-3 reactor is shown in FIG. 2

### III. PASSIVE SHIELDING OF ANTINEUTRINO DETECTOR AT THE SM-3 REACTOR

To carry out neutrino experiment at research reactor a detector has to be placed into passive shielding to protect it from background. In order to determine optimal parameters for the shielding the background conditions of the experiment were minutely investigated. Descriptions and results are presented in the following sections.

In order to fulfill requirements of the experiment and bring background conditions to acceptable level the passive shielding ("cabin") was constructed and its image from the outside and inside is shown in FIG. 3. The shielding is made of elements based on steel plates of size 1x2 m, 10 mm thick, to which are attached 6 lead sheets of 10 mm thickness. The cabin volume is 2x2x8 m. From the inside, the cabin is covered with plates of borated polyethylene of 16 cm thickness. The total weight of passive shielding is 60 tons, the volume of borated polyethylene is 10 m$^3$. Inside the passive shielding, there is a platform with an antineutrino detector, which can be moved along the rails within the range 6 - 12 meters from the center of the reactor core. The cabin (the neutrino beamline) can be entered by means of a ladder, through the roof with the removed upper unit, as shown in FIG. 3. The main hall of the reactor and our experimental hall are connected by a trapdoor in the celling of the hall. Loading of the detector into a neutrino beamline is carried out from the main hall through this trapdoor. In this case, an overhead crane of the main hall is used.

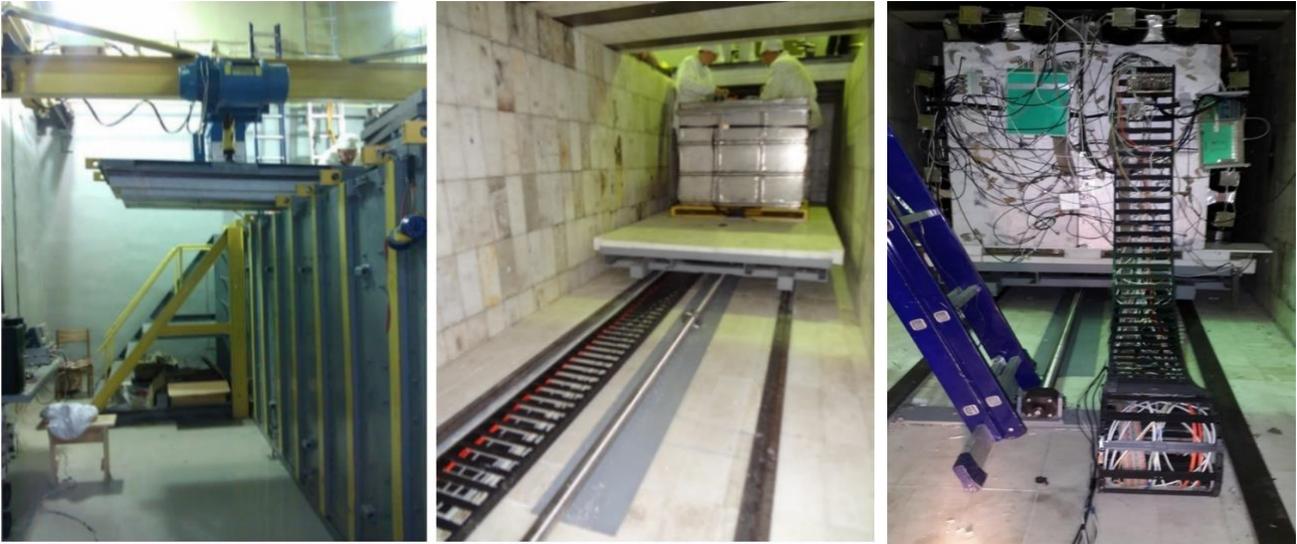

FIG. 3 General view of passive shielding: from the outside and inside. The range of detector movements is 6 - 12 m from the center of the reactor core on the left. Model detector is in the center, full-scale detector is on the right.

## IV. INVESTIGATION OF BACKGROUND CONDITIONS INSIDE AND OUTSIDE OF PASSIVE SHIELDING WITH A GAMMA DETECTOR

The detailed knowledge of background conditions around the detector are absolutely necessary to carry out neutrino experiment. We performed the detailed investigation of various types of backgrounds. Flux and spectrum of gamma particles was measured with a detector based on NaI(Tl) crystal of size 60×400 mm.

The first measurements of gamma background in the neutrino laboratory hall were carried out before installation of passive shielding. While the reactor was in operation mode, we registered gammas from neutron capture in an iron-concrete shielding of the reactor. During the reactor operation, the background of gamma radiation, in the energy range from 3 MeV to 8 MeV significantly (22 times larger than natural radiation background) increase, because of thermal neutrons interaction with iron shot contained in concrete shielding of the reactor. This energy range is of great importance, since it corresponds to energy of gamma-quanta emitted in the process of neutron capture by Gd, which we use to register antineutrinos in our detector.

Gamma radiation of isotopes $^{137}$Cs, $^{60}$Co is independent of the reactor operation mode and is caused by radioactive contamination from the building floor and walls. Concrete reinforced with iron grit was used for flooring and the slide valve was reconstructed. These modifications reduced in 5 - 6 times gamma radiation background in the energy range we are concerned with. Despite that, remaining gamma background was still too high, and it confirmed necessity of creation of passive shielding from gamma rays for the detector. Installation of passive shielding significantly suppressed gamma background from the reactor to the level of radioactive contamination in the passive shielding.

Within energy range of 1440÷7200 keV (from $^{40}$K line and higher), the 5 cm lead shielding makes the level of background gamma radiation 4.5 times lower, which proves that its installation on the detector is reasonable. However, fast neutron background, resulting from the interaction of cosmic muons with lead nuclei, enhances inside the lead shielding. Indeed, the 5 cm lead shielding around a neutron detector doubles its count rate. Therefore, a layer of borated polyethylene should be placed inside the lead shielding.

After installation of the passive shielding, we carried out detailed measurements of gamma background inside it to determine actual conditions around the detector. FIG. 4 presents the gamma spectrum inside the passive shielding for various distances along the route of the detector: 6.28 m, 8.38m, 10.48 m. No noticeable alterations in the spectrum shape were observed. Moreover, for comparison, gamma-spectra are measured at the reactor On and Off inside the passive shielding, at the point nearest to the reactor. Considerable difference in these spectra was not found.

## V. ESTIMATIONS OF FAST NEUTRON FLUX

In 2013, at the SM-3 reactor, all preparations of the neutrino laboratory room were completed as well as the installation of the passive shielding of the neutrino detector.

The slide valve of the former neutron beamline was carefully plugged. As a result, a flux of thermal neutrons in the neutrino laboratory room decreased 29 times to the level of $1 \div 2 \cdot 10^{-3} \; n/(cm^2 s)$. This level is an effect of cosmic rays and, practically, is independent of the reactor operation.

Measurements of thermal neutron flux were made with $^3$He detector, which is a proportional counter of 1 m long with diameter of 30 mm. For registration of fast neutrons, we used a same $^3$He detector, but it was placed into a shielding made of polyethylene (thickness of layer is 5 cm), which in its turn was wrapped in a layer of borated rubber (3 mm thick, containing 50% of boron). Thermal neutrons stop in borated rubber while fast neutrons penetrate it and slows down in polyethylene to be registered by $^3$He detector.

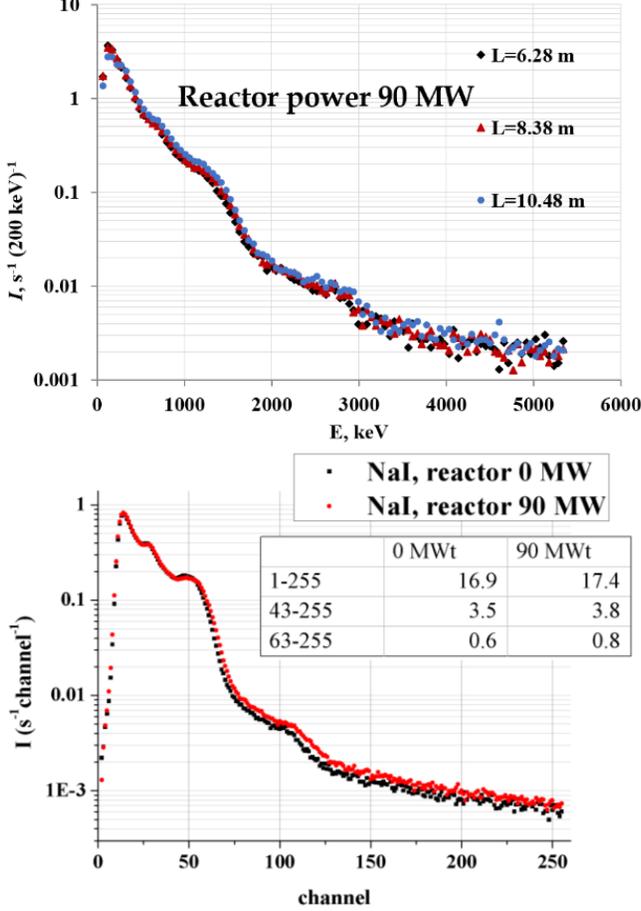

FIG. 4. Gamma spectra at the detector location. Top – reactor power is 90MW. L is distance from the center of the reactor core: - 6.28 m, - 8.38m, - 10.48m. bottom – reactor ON/OFF spectra.

To convert count rate (s$^{-1}$) of proportional counter $^3$He detector into neutron flux density (cm$^{-2}$·s$^{-1}$), both detectors were calibrated using standard detector MCS AT6102. For this purpose, $^3$He detector and MCS were placed side by side at the distance of 3 m from a neutron source (Pu-Be).

$^3$He detectors of thermal and fast neutrons have sensitivity two orders of magnitude higher than that of standard MCS detector. They were employed for conducting measurements of low background. Estimations of neutron background were made, for the first time, before upgrading former neutron beamline enter (before slide valve was plugged), then, after upgrading and, finally, after installing passive shielding of the neutrino detector.

In measurements of the fast neutron flux outside the passive shielding the detector of fast neutrons was located on the roof of the passive shielding near the reactor wall, i.e. at the distance of 5.1 m from the reactor core. The results of measurements are shown in FIG. 5 on the left. The flux is almost independent of the reactor power and its value is 10$^{-3}$ s$^{-1}$cm$^{-2}$. The excess above the level of the cosmic background at the full reactor power is (5±2)%. The measurements of the fast neutron flux inside the passive shielding were carried out in two modes. The first one was aimed on determining the influence of the reactor. For that purpose, the measurements of the fast neutron flux were performed inside the passive shielding at the position nearest to the reactor wall with operating reactor and reactor in off mode. Both measurements were carried out for 10 days. With operating reactor, the fast neutron flux was equal to (5.54±0.13) 10$^{-5}$ s$^{-1}$ cm$^{-2}$, while with the switched off reactor, it was (5.38±0.13) 10$^{-5}$ s$^{-1}$cm$^{-2}$, i.e. there was no difference within the accuracy of 2.5%, (0.16±0.13) 10$^{-5}$ s$^{-1}$cm$^{-2}$.

More accurate estimate of the fast neutron flux from the reactor within the passive shielding can be made using a suppression factor of 12 for the fast neutron flux inside the passive shielding. Then the excess of the fast neutron flux above the cosmic background level at the full reactor power inside the passive shielding is (0.38 ± 0.15) %. Thus, at a signal/background ratio of 0.5, the contribution of fast neutrons from the reactor to the ON-OFF neutrino signal can be (1.1 ± 0.45) %. This is not a problem for these measurements, especially since this background cannot have the oscillation behavior.

The second mode was aimed on direct measuring of fast neutron background inside passive shielding on the neutrino detector route. For that purpose, the detector of fast neutrons was installed on top of the neutrino detector and was moved with it inside the neutrino beamline in range from 6.25 m to 10.5 m from the reactor. The results of this measurements with operating and switched off reactor are shown in FIG. 5 (on the right). There is no difference caused by the reactor mode within statistical accuracy. Also the background does not depend on distance. In these measurements, the background level appeared to be equal to (8.5 ±0.1)10$^{-5}$ s$^{-1}$ cm$^{-2}$, which is somewhat higher than that near the reactor wall. The discrepancy can result from the detector positioning relative to direction of a neutron flux, near the reactor wall it was installed vertically, while on the top of the neutrino detector it was installed horizontally.

Finally, we can conclude that fast neutron background is almost independent of the reactor working mode, but it is determined by the neutrons created in interaction of cosmic rays muons with matter around the detector. Notice, that passive shielding contributes to neutron background, because muons interact with materials of the shielding. However, passive shielding suppresses the fast neutron background by the factor of 12, so that fast neutron flux outside the passive shielding is 10$^{-3}$ s$^{-1}$cm$^{-2}$, while flux inside the shielded area is (8.5 ± 0.1)10$^{-5}$ s$^{-1}$cm$^{-2}$.

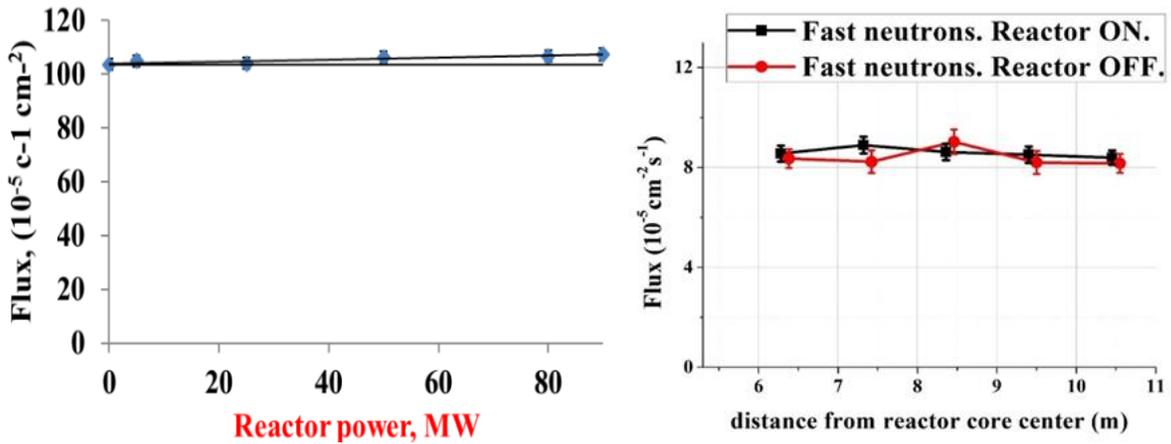

FIG. 5. Left – plot of neutron flux (near the reactor wall, i.e. at distance 5.1 m from the reactor core) as a function of reactor power. Right - Fast neutron background at various distances from the reactor core measured with the detector of fast neutrons inside passive shielding. The detector of fast neutrons was placed on top of the neutrino detector and was moved with it.

## VI. INVESTIGATION OF THE BACKGROUND CONDITIONS WITH ANTINEUTRINO DETECTOR MODEL

Before the measurements with full-scale neutrino detector we tried the model of it. The model of antineutrino detector contains 400 l of liquid scintillator BC-525 with addition of Gd with concentration 1 g/l, 16 PMTs on the top and 5 plates of active shielding (muon veto) around it. The model of the antineutrino detector and active shielding installed in the passive shielding is shown in FIG. 6.

FIG. 7 presents the spectrum obtained with the antineutrino detector model which can be divided into 4 parts. The first part (up to 2 MeV) corresponds to the background radioactive contamination; the second part (from 2 MeV to 10 MeV) covers the neutron/neutrino registration area, since it corresponds to energy of gammas emitted in neutron capture by Gd. The range from 10 to 60 MeV is related to soft component of cosmic rays which is a result of muon decays and muon captures in matter. And finally, the range 60 – 120 MeV is related to muons passing through the detector. This picture also illustrates small alterations of the spectrum shape for different detector positions.

In the course of long-term measurements [24,25], temporal variations of cosmic ray intensity have been found. They are caused by fluctuations of atmospheric pressure and season changes of temperature. These are well-known barometric and temperature effects of cosmic rays (FIG. 8 and FIG. 9)

Behavior of fast and slow components differs by additional long-term drift, with the drift sign being opposite for fast and slow components. It is the so-called temperature effect which can be interpreted in the following way. At the rise of the temperature in lower atmospheric layers, their expansion results in increase of the average altitude of creation of muon fluxes. As the distance to the Earth grows, the share of the decayed muons is getting larger. Thus, the intensity of fast component (muons) decreases and that of slow component (products of decay: electrons, positrons, gamma quanta) rises.

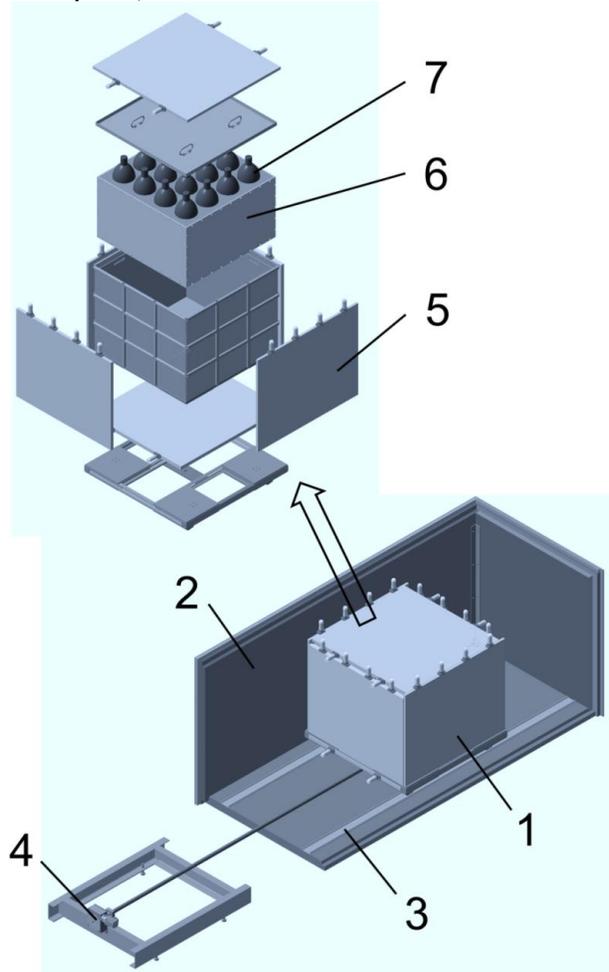

FIG. 6. Model of the neutrino detector installed in the passive shielding [24,25]. 1 – detector of reactor antineutrino, 2 – passive shielding, 3 – rails, 4 – engine for detector movement, 5 – active shielding with PMT, 6 – volume with liquid scintillator with Gd, 7 – detector's PMT.

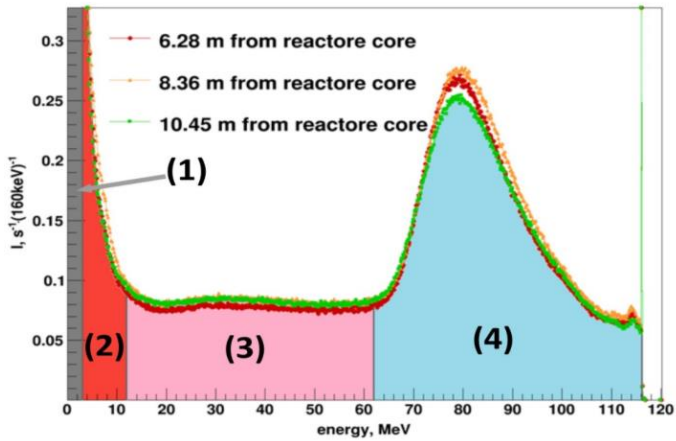

FIG. 7. Detector energy spectrum at different distances from the reactor core and a division of spectrum into zones: 1 – radioactive contamination background, 2 - neutrons, 3 –soft component of cosmic rays, 4 -muons.

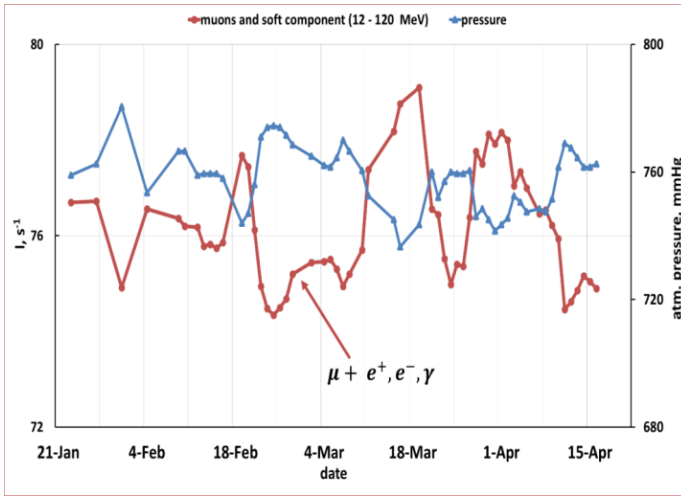

FIG. 8. Barometric effect of cosmic rays: the left axis illustrates a summary detector count rate in the energy areas 3 and 4, the right axis shows atmospheric pressure, the horizontal axis gives the measurement time since 23 of January till 15 of April in 2014.

FIG. 9 illustrates the drift effect with opposite signs for fast and slow components of cosmic background with increasing temperature of the lower layers of atmosphere in the vicinity of the Earth surface since January till April: from – 30 C to +10 C.

The fluctuations in the cosmic background are determined by the fluctuations in atmospheric pressure that is about ±1.1%. From the studies of variation in the cosmic background, an important quantitative conclusion can be made about their effect on measurements.

The direct results of variation in the cosmic background of the statistical distribution of the neutrino signal will be presented in section XII in FIG. 31.

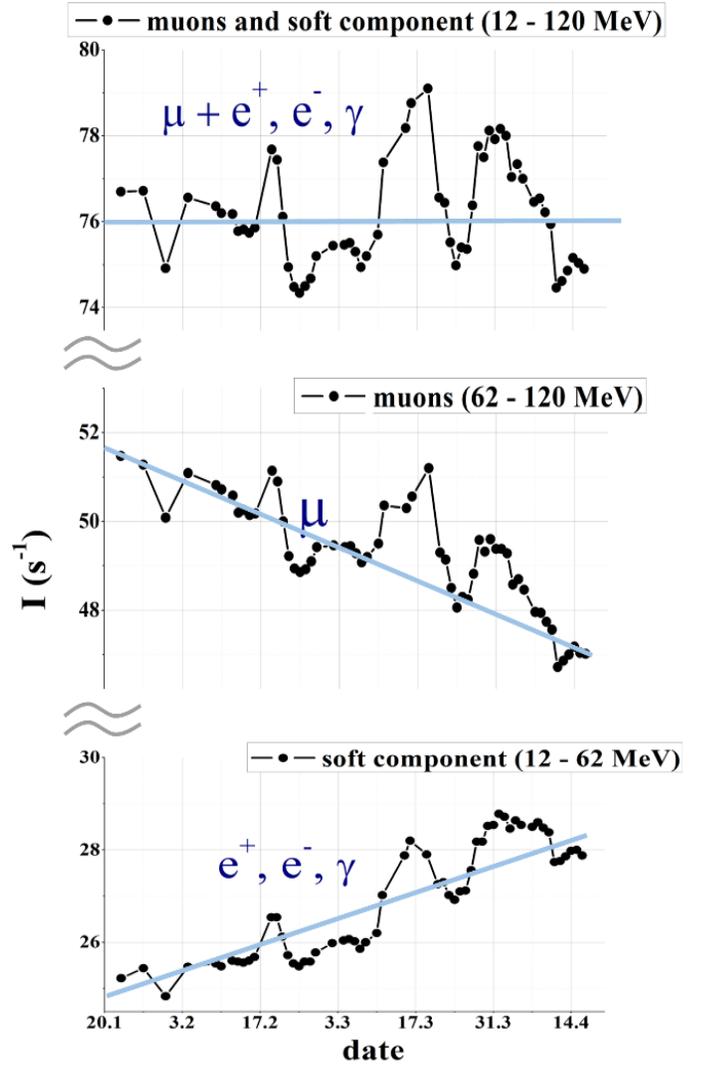

FIG. 9. The barometric and temperature effects of cosmic rays: top – a summary detector count rate in the energy areas 3 and 4, middle – a detector count rate in the energy area 4, bottom – a detector count rate in the energy area 3. The horizontal axis is the measurement time since 23 January till 15 April of 2014.

## VII. COSMIC BACKGROUND, ACTIVE SHIELDING. ENERGY AND TIME SPECTRA OF CORRELATED SIGNALS

The antineutrino flux is measured with inverse beta-decay reaction $\bar{\nu}_e + p \rightarrow e^+ + n$. For the registration of these signals we use delayed coincidence method.

In that measurement scheme an event starts with a registration of a signal with suitable characteristics. It opens a time window in which we expect to register a delayed signal with another set of suitable parameters. In the end of the time window when the probability of neutron capture is very low the background of accidental coincidence is measured. The measured value of the accidental coincidence background can be subtracted from the results.

In our experiment the correlated background is caused by cosmic ray muons. Therefore, when the reactor is switched off our detector and the method of delayed coincidence can be used to measure muon background. Muons in the detector create a delayed signal if either muon stops in the scintillator creating a muon atom where this muon decays with lifetime of 2.2 $\mu s$ or muon interaction with matter inside the detector results in emission of evaporation neutron which is captured by gadolinium after slowing down for 5$\mu s$. The characteristic time of muon capture by Gd in scintillator with Gd concentration of 0.1% is 31.3 $\mu s$.

FIG. 10 illustrates time spectra of delayed coincidences. The background of accidental coincidences is subtracted. The upper black curve represents measurements without using of active shielding of the detector. One can see two exponents (straight lines in logarithmic scale), which correspond to a muon decay and a neutron capture by Gd. The integral under the first exponent corresponds to stopped muons count rate of 1.54 µ/s, and the slope corresponds to a muon lifetime (2.2 µs). The integral under the second exponent corresponds to a neutron capture rate in the detector – 0.15 s$^{-1}$, and the parameter of the exponent (31.3 µs) corresponds to the neutron lifetime in the scintillator with 0.1% of Gd.

The number of muon stops per second agrees with the estimation based on data about muon flux and the scintillator mass, while neutron capture rate agrees with the calculated rate of neutrons generated in the detector itself, as a result of the muon flux passing through it. Adequacy of the installed passive shielding is confirmed by the fact that additional 10 cm of borated polyethylene above the neutrino detector do not change fast neutron count rate.

Muon background can be significantly suppressed by employing of active shielding of the detector and rejecting detector signals with too high energy. If the system gets the signal from active shielding or if measured energy in the detector exceeds 9MeV then the system stops registering signals for time 100 µs.

One of the major problems in our experiment is to separate correlated events from background of accidental coincidences. An example of measurements of spectrum of delayed signals is shown in FIG. 11. The lifetime of neutron in the scintillator with Gd is 31.3 $\mu s$, so a neutron will be captured in 200 $\mu s$ after the prompt signal with probability higher than 99%. Assuming that background of accidental coincidences has uniform distribution in time we can use an interval after 200 $\mu s$ to measure it. So, measuring rate of delayed coincidences we also control rate of background of accidental coincidences at the same time. To measure background, we chose interval of 100 $\mu s$ so the total time window in which we expect a delayed signal is 300 $\mu s$.

Besides the time in which a delayed signal occurs we have another parameter to select neutrino events the energies of prompt and delayed signals. When we determine exact energy region in which we search for signals we try to obtain the best signal to background ratio. A positron signal has natural threshold of 1 MeV the total energy of its annihilation. Therefore, lower limit of registered energy range is higher than 1 MeV. The lower limit is the more antineutrino events we register but at the same time the higher is background of accidental coincidences. Time and energy spectrum of delayed signals obtained with threshold 3 MeV are shown in FIG. 11.

The background of accidental coincidences is sufficiently low while the amount of antineutrino events is at the acceptable level. The point is a signal of neutron capture by Gd has sufficient amount of energy – up to 8Mev, while background of natural radioactivity is negligible on energy region above 3MeV. In decreasing of lower threshold from 3 MeV to 1.5 MeV the rate of accidental coincidences considerably increases (FIG. 12).

The upper energy limit is determined by spectrum of reactor antineutrino and emission of energy in neutron capture by Gd. The lower limit in the neutrino signal registration must be 1 MeV. The study of influence of energy limits on background and efficiency of antineutrino registration was concluded in setting ranges for prompt and delayed signals 1.5 – 9 MeV and 3 – 12 MeV correspondingly. Using the active shielding veto and selecting signals by energy we managed to suppress the background of correlated signals caused by neutrons to the level of $1.1 \cdot 10^{-2}$ $s^{-1}$. We assume that the remaining background is caused by fast neutrons, which appears in interactions of cosmic muons with matter outside the detector. In that case the prompt signal is created by a recoil proton. That background cannot be suppressed by using the active shielding because it does not register fast neutrons. The background of fast neutrons emitted outside the detector appears to be the main problem of our experiment and the solution of this problem is described in the next section.

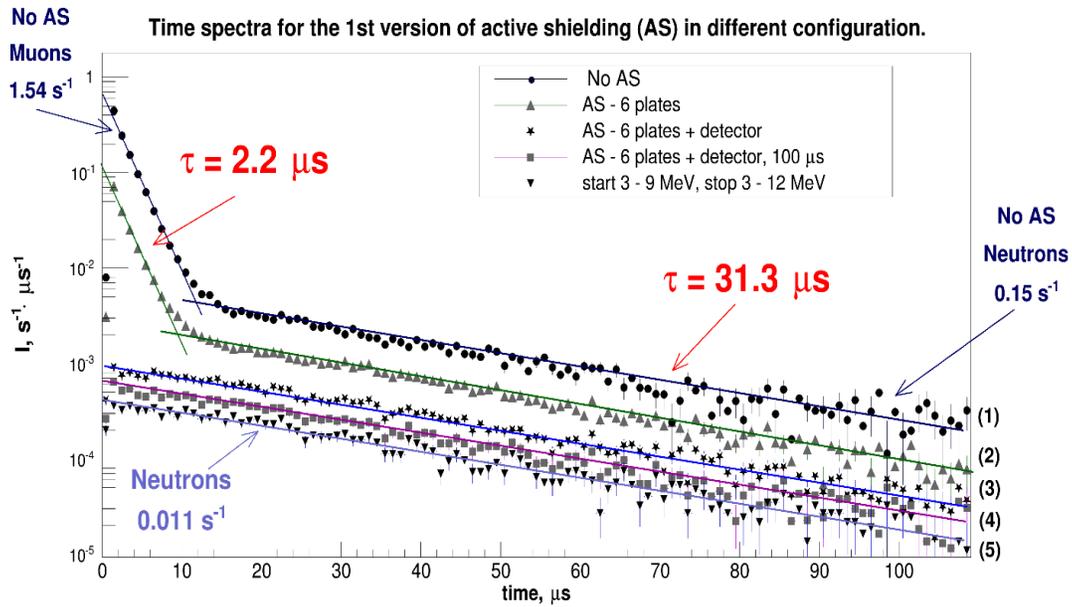

FIG. 10. Time spectra at different configurations of active shielding: 1 - no active shielding, 2 – plates of active shielding are using, 3 – the same + ban from the detector at signals higher than 12 MeV, 4 – the same + ban on 100 µs after the detector signal, at energy higher than 12 MeV, or after the signal in active shielding, 5 – the same + limit on start and stop signals in ranges 3 – 9 MeV and 3 – 12 MeV respectively.

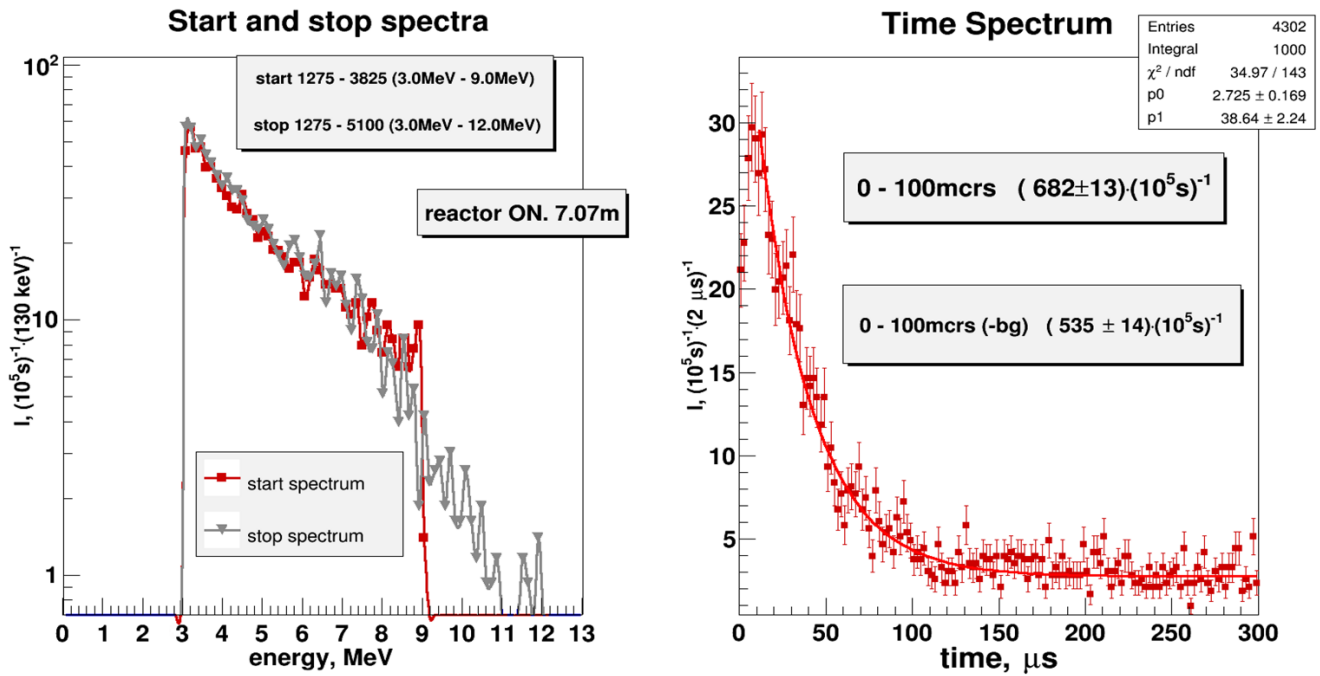

FIG. 11. Energy spectra of prompt and delayed signals (left) and time spectra (right): threshold of start and stop signals 3 – 9 MeV and 3 – 12 MeV correspondingly.

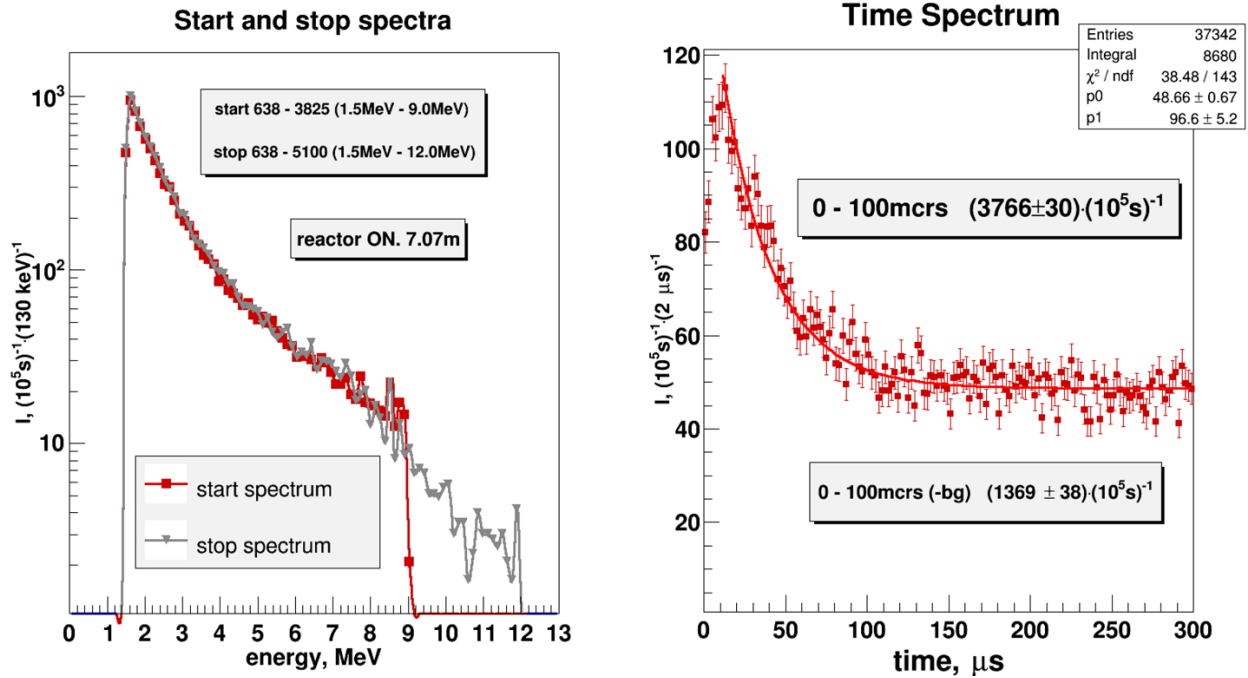

FIG. 12. Energy spectra of prompt and delayed signals and time spectra: 1.5 – 9 MeV and 1.5 –12 MeV.

## VIII. CARRYING OUT RESEARCH WITH A MODEL OF ANTINEUTRINO DETECTOR OF A MULTI-SECTION TYPE

The first measurements carried out with the detector model revealed that combination of the active and passive shielding and selection of events by energy are yet not enough to suppress the correlated background. After suppressing other sources of background events, fast neutrons emitted outside the detector in interactions of high energy cosmic ray muons with matter around the detector become the main source of the background.

The scattering of fast neutrons easily imitates an IBD process, which we use to detect neutrinos. Registration of the first (start or prompt) signal from recoil protons imitates registration of a positron. The second (stop of delayed) signal arises in both cases when a neutron is captured by gadolinium. The active shielding cannot help to distinguish fast neutron signals from antineutrino signals (FIG. 13).

Multi-section scheme was developed to get additional selection criteria for antineutrino events. There is a difference in localization of prompt signals of antineutrino and neutron events. A recoil proton in matter has track of about 1 mm length, while a positron emitted in IBD process annihilates with emission of two gamma-quanta each having energy 511 keV and opposite directions. As a result, if the vessel with the scintillator is divided in several sections of the same size with walls reflecting optical photons, then the track of recoil proton will be contained within one section. The track of a positron has average length of about 5 cm, so its signal is also registered in one section, but gammas with energy 511 keV can be registered in adjacent sections.

The detector inner vessel was divided into 16 sections 0.225 x 0.225 x 0.5 m³ with rigidly fixed partitions between them. At the same time, we started to use the active shielding consisting of two layers external and internal relative to passive shielding. The external layer ("umbrella") moves on the roof of the passive following the detector movements [26]. The scheme of locations of the multi-section detector and active shielding relative to the passive shielding is shown in FIG. 14.

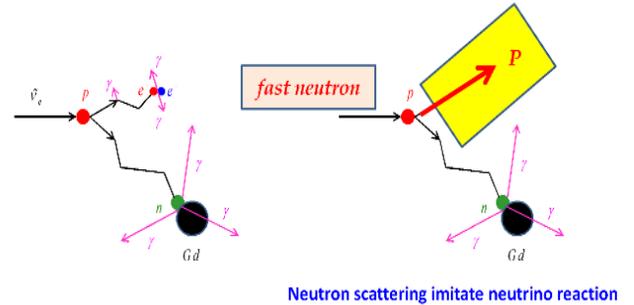

FIG. 13. Illustration of the background problem caused by the fast neutrons.

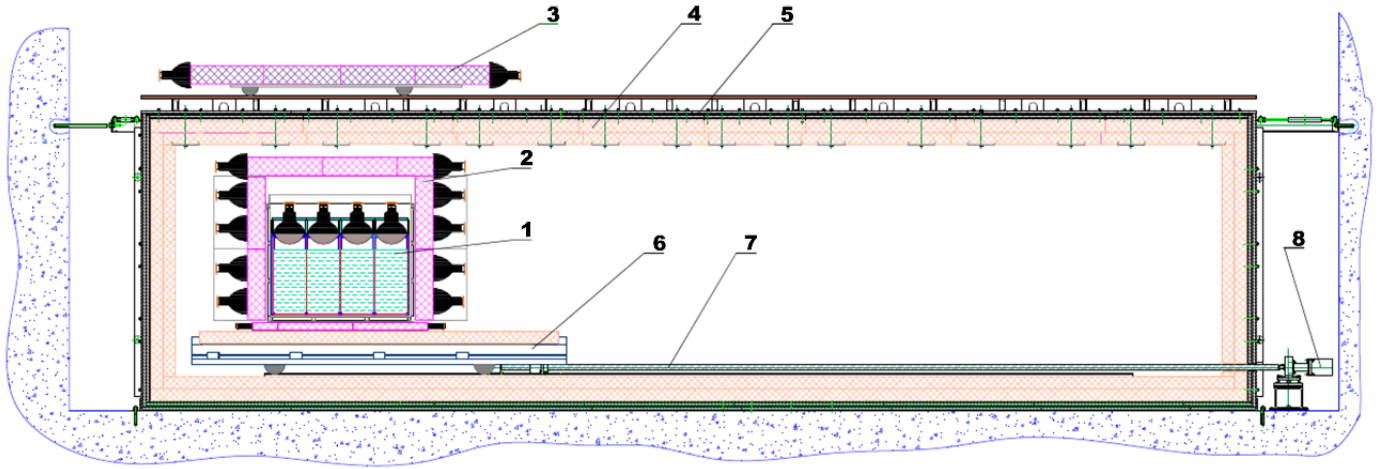

FIG. 14. General scheme of an experimental setup: 1 – detector of reactor antineutrino, 2 – internal active shielding, 3 – external active shielding (umbrella), 4 – borated polyethylene passive shielding, 5 – steel and lead passive shielding, 6 – moveable platform, 7 – feed screw, 8 – step motor.

The ratio of single-section and multi-section prompt signals of antineutrino events depend on detector configuration, amount of sections and their sizes. Therefore, a Monte-Carlo simulation of particular detector configuration is required.

The detector scheme for Monte Carlo calculation is presented in FIG. 15. Probability of recording double starts depends on the section location: in the center, on the sides or in the corner. The probability of registration of double starts for different sections where the event occurred is presented in Table I.

TABLE I. Probability of registration of double starts.

| central cell | side cell | angular cell | in all cells |
|---|---|---|---|
| 0.424 | 0.294 | 0.188 | 0.300 |

The mean probability of double starts over all detector is 30%. That means in our method 70% of prompt signals of antineutrino events occurs in single section. Therefore, if we only consider events with double starts then the number of registered events decrease in 3 times which is obviously unacceptable. But the analysis of results can be performed by using the single-section and multi-section events, and use their ratio to the total amount of the events (30% and 70%) as an additional criterion which can be used to check the validity of selection of antineutrino events. Thus, if signal difference between the reactor ON and reactor OFF measurements is within 30% - 70% for the multi-section and single-section events, then it can be interpreted as a neutrino signal.

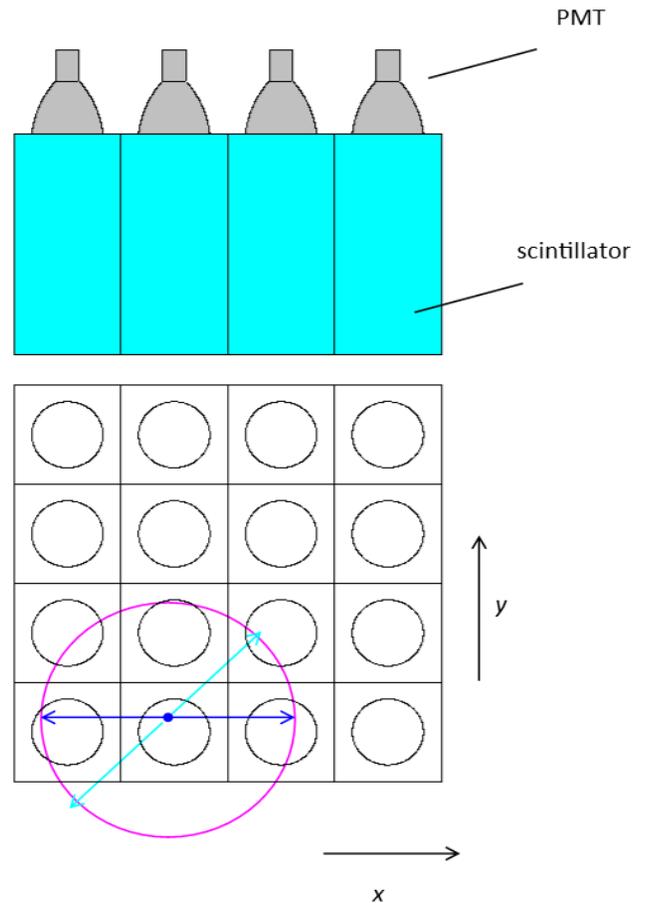

FIG. 15. Scheme of the detector of reactor antineutrinos.

Preliminary measurements with the Pu-Be fast neutron source have been made before the start of measurements with the new multi-section detector model. The time spectra of the single-section and multi-section prompt signals are shown in FIG. 16. It appears, that if we consider only multi-section prompt signals, then the correlated signals from neutrons are completely excluded and only a straight line from an accidental coincidence remains. This experiment has revealed that fast neutrons give only single-section starts.

The count rate difference (ON-OFF, i.e. with the reactor switched on and off) for two-section and single-section starts, integrated over all distances, makes up (29±7) % and (71±13) % respectively. Within the available accuracy, such a ratio allows to assume the registered events as neutrino-like events. Thus, it is another evidence that ON-OFF signal corresponds to neutrino events.

However, the accuracy of this statement is lower than direct estimates of the contribution of fast neutrons from the reactor to the ON-OFF neutrino signal, which can be (1.1 ±0.45)%, as shown in section V.

Unfortunately, we cannot use the two-section starts selection, since we lose significantly in data collection.

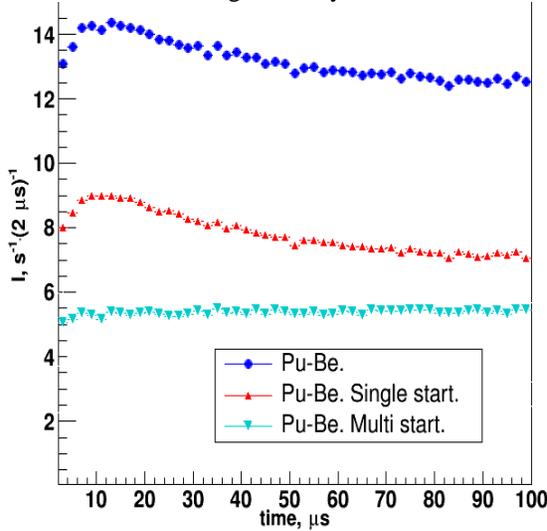

FIG. 16. Time spectra of the delayed coincidence obtained with a fast neutron source. The red curve corresponds to single section starts, and the green one shows multi-section starts.

In order to determine the energy resolution of the detector a single section detector was manufactured to carry out detailed research. We use the effect of total internal reflection of the light from the border between scintillator and air at low angles to align collection of light (make it homogeneous) from various distances. The problem is that in case of optical contact of scintillator and PMT the light is better collected from the distances close to PMT in solid angle close to $2\pi$. The light from farther positions comes through the mirror light guide, while efficiency of light transportation for angles close to the right angle is much worse because of multiple refractions. The effect of total internal reflection at the border of scintillator and air at low angles of descent evens the transportation conditions for light coming from various distances. Finally, a mirror at the bottom of the light guide also helps to even light collection conditions for various positions in the detector section. Demonstration of the effects described above is shown in FIG. 17 on top.

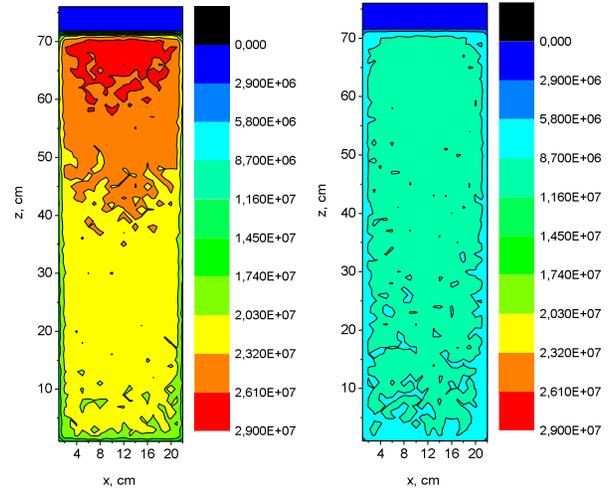

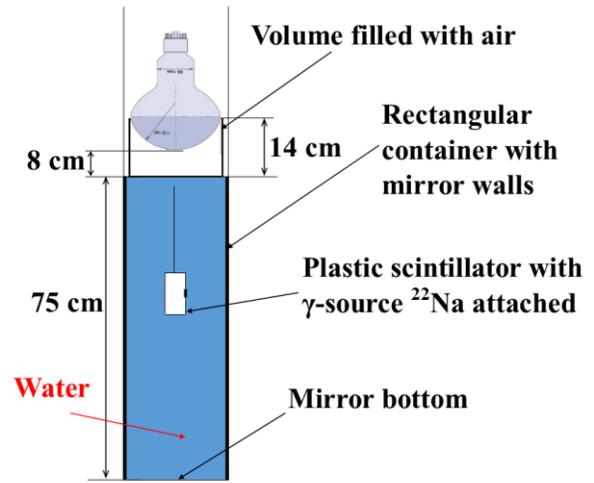

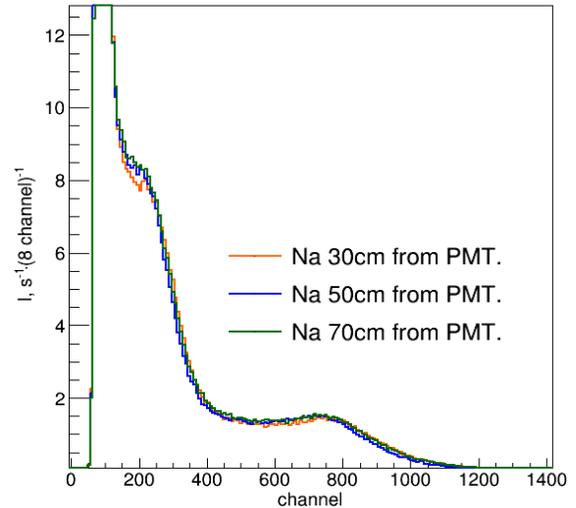

FIG. 17 Top – results of MC simulation for model of single section in case with optical contact (on the left) and without optical contact (on the right). Middle – scheme of model to measure with full-scale detector section analogue. Bottom – $^{22}$Na source spectrum with different scintillator position for model of full-scale detector section with air gap.

The scheme of the experiment with single section is shown in FIG. 17 in the middle. To carry out research of efficiency of light collection with usage of the total internal reflection the section was filled with water which has the refraction index close to the index of the scintillator. To simulate scintillation, we used a source made of plastic scintillator and a gamma source $^{22}$Na with lines 511 keV and 1274 keV. The location of scintillation was determined by location of the source. As one can see in FIG. 17 (bottom) the gamma lines are almost independent of the location of the source 30 cm, 50 cm and 70 cm from the water surface. Therefore, the calibration of the detector with scintillator can be carried out with a source outside the detector. That fact is very convenient for performing the calibration procedure.

## IX. THE FULL-SCALE ANTINEUTRINO DETECTOR

The model of the detector was replaced with the full-scale detector in 2016. This detector is also filled with liquid scintillator with gadolinium concentration 0.1%. The detector inner vessel is divided into 50 sections – ten rows with 5 sections in each having size of 0.225x0.225x0.85m$^3$, filled with scintillator to the height of 70 cm. The total volume of the scintillator is 1.8 m$^3$. The detector is placed into the passive shielding. The neutrino detector active shielding consists of external and internal parts relative to passive shielding. The internal active shielding is located on the top of the detector and under it. The scheme of the detector and shielding is shown in FIG. 18.

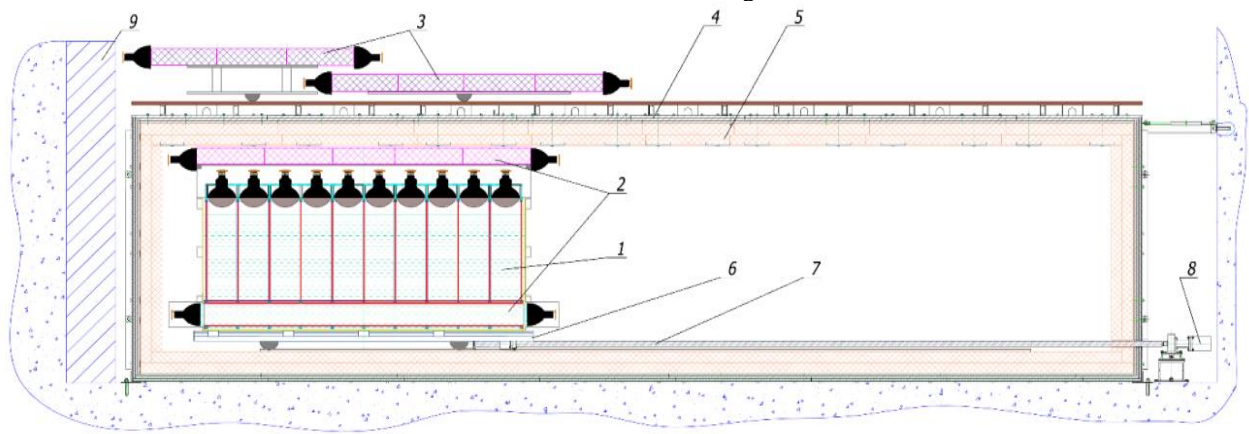

FIG. 18. General scheme of an experimental setup. 1 – detector of reactor antineutrino, 2 – internal active shielding, 3 – external active shielding (umbrella), 4 – steel and lead passive shielding, 5 – borated polyethylene passive shielding, 6 – moveable platform, 7 – feed screw, 8 – step motor, 9 –shielding against fast neutrons made of iron shot.

The first and last detector rows were also used as an active shielding and at the same time as a passive shielding from the fast neutrons. Thus, the fiducial volume of the scintillator is 1.42 m$^3$. For carrying out measurements, the detector has been moved to various positions at the distances divisible by section size. As a result, different sections can be placed at the same coordinates with respect to the reactor except for the edges at closest and farthest positions.

Monte Carlo calculations has shown that 63% of prompt signals from neutrino events are recorded within one section and only 37% of events create signal in an adjacent section [26]. In our measurements, the signal difference at the reactor ON and OFF has ratio of double and single prompt events integrated over all distances $(37 \pm 4)$ % and $(63 \pm 7)$ %. This ratio allows us to interpret the recorded events as neutrino events within current experimental accuracy. Unfortunately, a more detailed analysis of that ratio cannot be performed due to low statistical accuracy. Yet, it should be noted, that the measurements of fast neutrons and gamma fluxes dependence on distance and reactor power were made before installing the detector into passive shielding (Sec. V and [24,25]). Absence of the noticeable dependence of the background on both distance and reactor power was observed. As a result, we consider that difference in reactor ON/OFF signals appears mostly due to antineutrino flux from the operating reactor. That hypothesis is confirmed by the given above ratio of single and multi-section prompt signals typical especially for neutrino events.

The measurements of fast neutrons and gamma fluxes in dependence on distance and reactor power were made before installing the detector into passive shielding. Absence of noticeable dependence of the background on both distance and reactor power was observed. As a result, we consider that difference in signals (reactor ON - reactor OFF) appears mostly due to antineutrino flux from operating reactor. The signal generated by fast neutrons from reactor does not exceed 3% of the neutrino signal. The fast neutron background is formed by cosmic rays. The averaged over distance ratio of ON-OFF (antineutrino) signals to background is 0.5

## X. ENERGY CALIBRATION OF THE DETECTOR

Properties of one section were investigated earlier in section VIII. It was revealed that the energy resolution is independent of the position of event registration. Therefore, detector calibration can be performed with source outside the detector on top of it (see FIG. 19)

Energy calibration of the detector was performed with γ-quanta source and neutron source ($^{22}$Na by lines 511 keV and 1274 keV, by line 2.2 MeV from reaction np-dγ, by gamma line 4.44 MeV from Pb-Be source, and also by total energy of gamma quanta 8 MeV from neutron capture in Gd) [27]. These calibration spectra are shown in FIG. 20 and more detailed in FIG. 21-22.

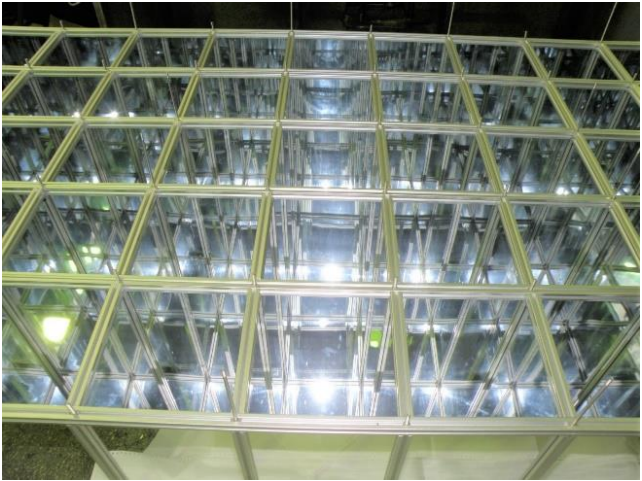

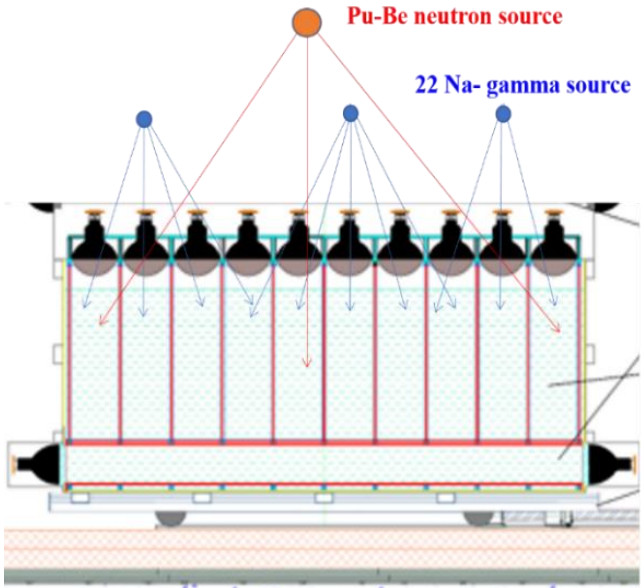

FIG. 19. The structure of the detector (top). Scheme of detector energy calibration (bottom).

FIG. 23 demonstrates linearity of calibration dependence. As a result, the spectrum of prompt signals registered by detector was measured. Its connection with antineutrino energy is determined by equation: $E_{promt} = E_{\bar{\nu}} - 1.8 \text{ MeV} + 2 \cdot 0.511 \text{ MeV}$, where $E_{\bar{\nu}}$ - is antineutrino energy, 1.8MeV – energy threshold of IBD, and $2 \cdot 0.511$ MeV corresponds to annihilation energy of a positron.

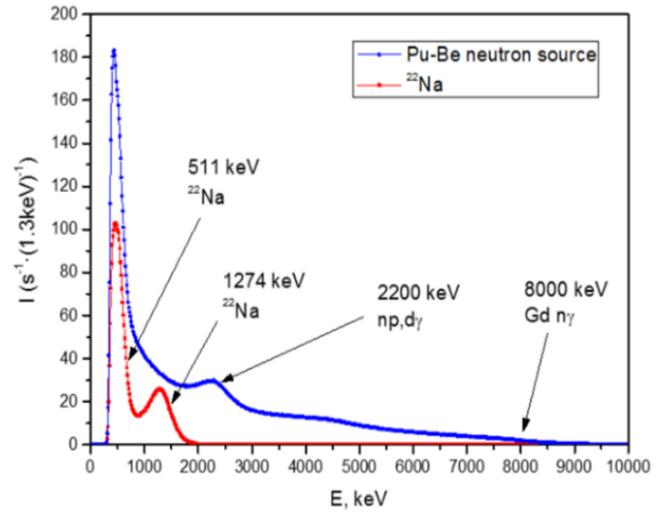

FIG. 20 The results of detector calibration.

The calibration above 2 MeV is a problem because of absence of required gamma sources in this energy region. Therefore the energy resolution in FIG. 23 is extrapolated assuming square root dependence from energy. The calibration demonstrated above and energy resolution correspond to a single section.

Actually, we are interested in energy resolution of the detector consisted of several sections. When a signal in a section is registered signals from closest sections are considered too. Therefore, energy resolution of the detector is better than energy resolution of one separate section.

Registration of a positron is more complicated process than gamma-quanta registration. The track of a positron has average length of about 5 cm, so its signal likely to be registered in one section. While gamma-quanta often leave the section still having significant amount of energy. Therefore, positron detection could be with higher energy resolution than detection of gamma-quanta. Unfortunately, positron registration is accompanied with emission of two gammas with energies 511 keV which can be registered in adjacent sections. It is important to notice that registration of the gammas leads to energy shift and broadening of the spectrum which are the same for all positron energies. Thus, it is incorrect to use energy calibration for single section with gamma peaks corresponding to sources and extrapolation of resolution assuming square root dependence for positron energy reconstruction. Further we will return to discussion of this problem because energy resolution of positron registration is extremely important for description of the oscillation process.

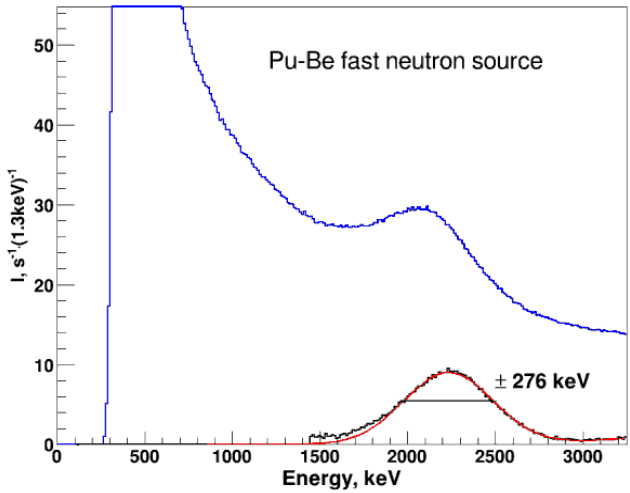
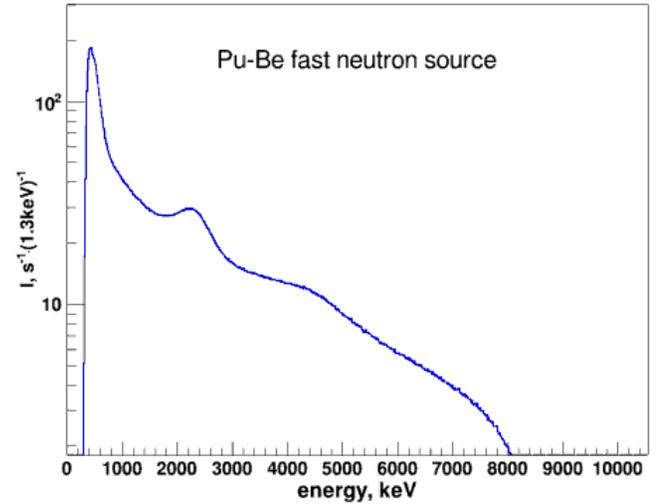

FIG. 21. The results of detector calibration. Left – line 2.2Mev of np-dγ process; right – 8 MeV from Gd(n,γ).

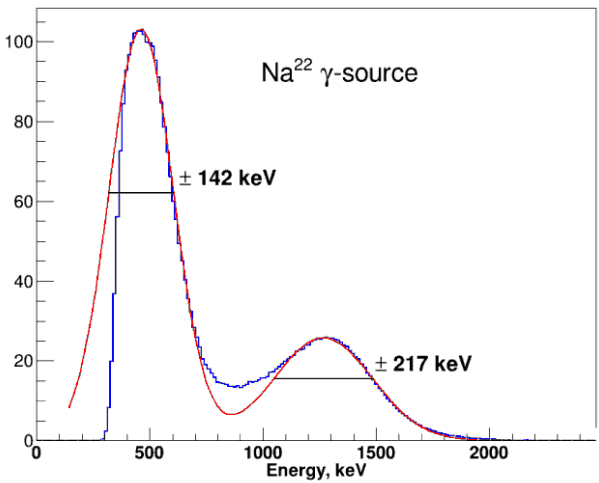

FIG. 22. Calibration with $^{22}$Na source.

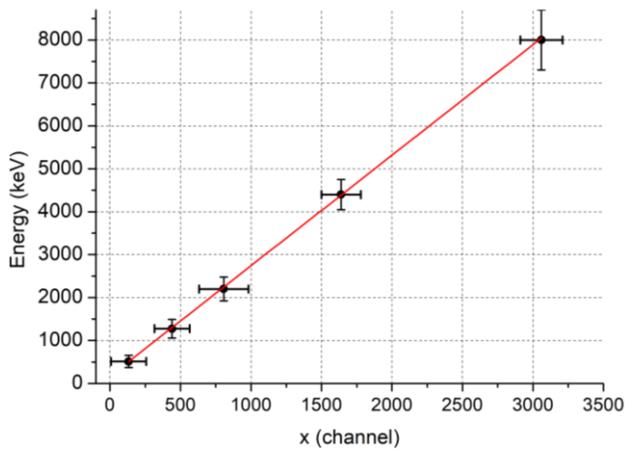

FIG. 23. Linearity of energy calibration.

## XI. COMPUTER MODEL OF REACTOR ANTINEUTRINO DETECTOR

In order to estimate the efficiency of antineutrino detector a computer model of the detector was created and using it Monte-Carlo calculation were carried out. Size of the detector and properties of the IBD process were used as the model parameters. In MC calculations in annihilation of stopped positron two gamma-quanta are emitted with energies 511 keV propagating in the opposite directions. The neutrons from the IBD process are absorbed by Gd with emission of gamma cascade of total energy 8 MeV. The detector register two successive signals from positron and neutron. The detector scheme used in the model is shown in FIG. 24.

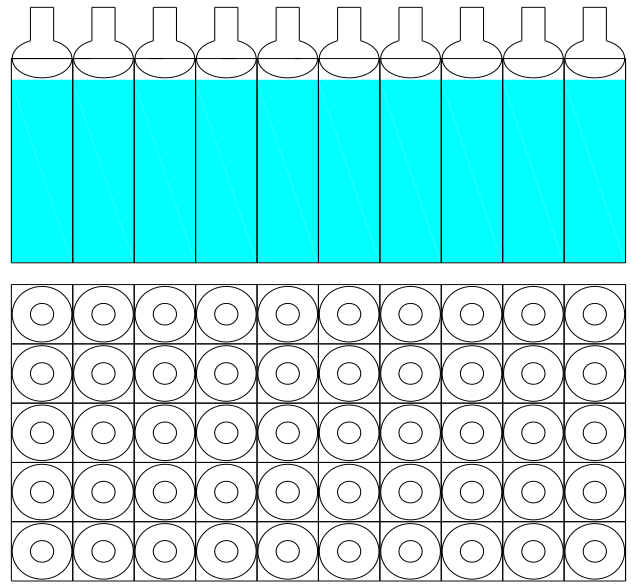

FIG. 24. Scheme of the detector of reactor antineutrino.

The detector vessel is divided into 5x10 sections 0.225 x 0.225 x 0.75 m³ with rigidly fixed partitions between them. Scintillator material is mineral oil ($CH_2$) doped with Gd of concentration 1 g/l. Scintillator light yield is $10^4$ photons per 1 MeV. Thickness of walls was neglected. Hamamatsu R5912 are used in the model. A layer of air separates PMT from scintillator. Antineutrino spectrum is calculated from positron spectrum, because to the first order approximation is can be represented as a linear function: $E_{\bar{\nu}} = E_{e^+} + 1.8$ MeV.

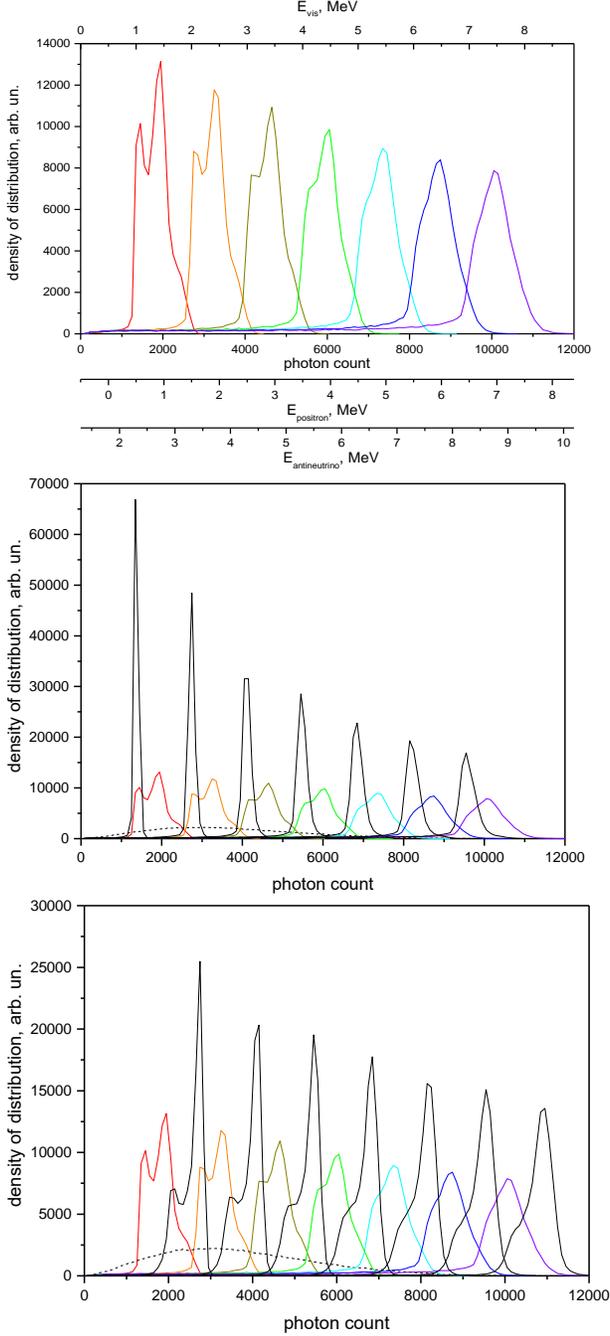

FIG. 25. Distribution of counts of PMT in one section induced by positrons with energies from 1 to 7 MeV with annihilation process (2 gamma-quanta with energies 511keV).

To simulate antineutrino spectrum we used antineutrino specrum of $^{235}U$[30]. In the model antineutrino flux has direction parallel to detector axis. This aproximation is valid for detector distances in range 6 – 12 m from the reactor core. Positron directions have isotropic distribution. Space distribution of neutron captures by Gd and energy yields of positron and gammas in scintillator were calculated with MCNP program[33]. The gamma spectrum of neutron capture by Gd was generated using spectrum of process $^{157}Gd(n,\gamma)$. Exponential track length of a photon into the scintillator is 4 m. Probability of photon reflection from the wall is 0.95.

A distribution of PMT counts (number of registered photons) from positrons of various energies and two gamma-quanta with energies 511keV is shown in FIG. 25. The top image shows the distribution of signals for one section taking into account the signals from the registration of two gammas with energies 511keV. It is mostly determined by incomplete absorption of gamma-quanta within one section. It explains two peaks in distribution at low energies. The middle image illustrates the distribution of signals from positron only without two gamma-quanta from the annihilation process. The distributions from the top image are also present at the middle image for comparison. The bottom image demonstrates the signals taking into account the annihilation gammas and registration in adjacent sections and distributions from the top image for comparison. This distribution is asymmetrical and hence it is difficult to estimate corresponding energy resolution. However, one can take the effective $2\sigma$ distribution width, which covers 68% of the total area under the curve. The calculations (table II) reveal that the effective width obtained from the results presented at bottom image almost does not depend on energy of positrons and can be estimated as $2\sigma = 500$keV.

TABLE II

| $E_{pos}$, MeV | $\sigma$, keV |
|---|---|
| 2 | 215 |
| 4 | 233 |
| 6 | 251 |

Examples of registration of gamma quanta with energies 2.3 MeV 4.4 MeV 6.0 MeV are shown in FIG. 26 for the case of registration by only one section (green curve) and the case of registration taking into account signals from nearest sections (red curve). The energy resolution in case of registration in several sections is two times better $2\sigma = 250$ keV for positron energy 4.4 MeV. The presence of partitions between adjacent sections should decrease the energy resolution. We will return to the question of the energy resolution of the detector in section XIV.

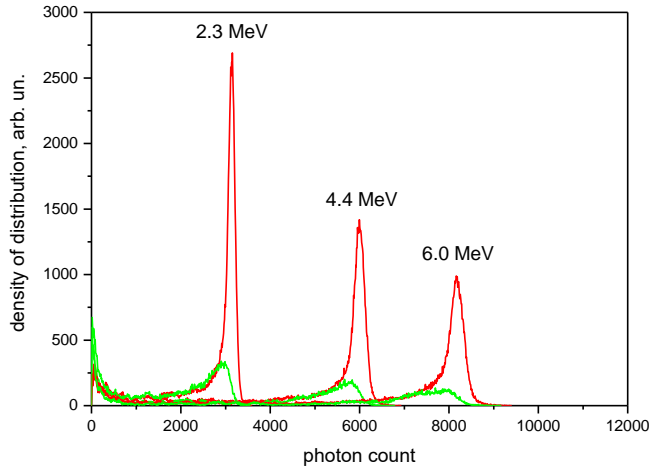

FIG. 26. Distribution of counts of PMT in one section induced by gamma quanta with energies 2.3 MeV 4.4 MeV 6.0 MeV

Distributions of signals of positron and neutron events are shown in FIG. 27. The threshold of positron event is 1.5 MeV. The efficiency of positron registration obtained with taking into consideration the positron spectrum and threshold is $\epsilon_{e+}=0.810(5)$.

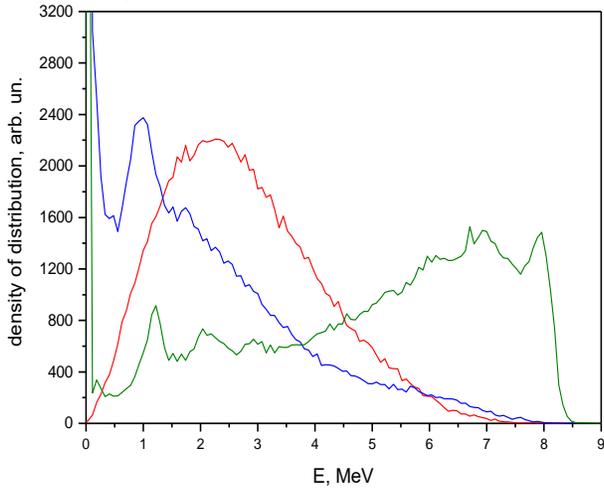

FIG. 27. Signals of positon (positron and 2 gamma-quanta) event - red curve, signals of neutron event – blue curve (only PMT of section where the IBD process took place is used), green curve – signals of PMTs from all sections.

The lower limit of energy in registration of neutron signal is planned to be set at level 3.2MeV, which will help to suppress the accidental coincidence background caused by natural radioactivity[25]. With such limit the registraton efficiency of neutron signal from $^{157}$Gd(n,γ) is $\epsilon_n=0.713(5)$ if we consider counts of PMTs of all sections. If we consider counts of PMT in one secton where the process $^{157}$Gd(n,γ) takes place $\epsilon_n=0.194(5)$. Taking into account the fact that ~20% of neutrons are captured by hydrogen with energy yield 2.2 MeV the efficiency is $\epsilon_n=0.570(5)$. The registration efficiency of IBD obtained in simulation is $\epsilon=0.462(5)$. If consider only PMT in section where the process occurred, then $\epsilon=0.128(5)$. Efficiency of the detector as function of limits of positron and neutron signals with counts of all PMTs is shown in FIG. 28.

In the experiment the limits of positron and neutron signals are set to be 1.5 MeV and 3.2 MeV correspondingly. For that values MC simulation gives detector efficiency ~46%. If we consider counts from PMTs of all sections, also consider influence of materials in scintillator volume and incomplete signal collection from process $^{157}$Gd(n,γ) in different sections then detector efficiency can be estimated to be ~20%.

The ratio of the expected neutrino count to the register is 25-30%, which is explained by that estimates.

It should be noted that accurate estimate of the detector efficiency is not the task of this experiment, since we use the method of relative measurements through the use of a mobile detector.

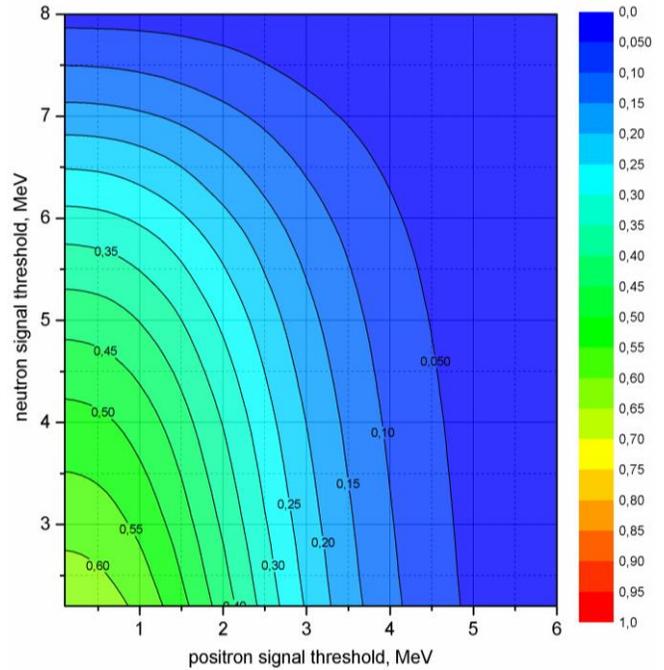

FIG. 28. Efficiency of the detector as function of limits of positron and neutron signals with counts of all PMTs.

## XII. MEASUREMENTS – THE SCHEME OF REACTOR OPERATION AND DETECTOR MOVEMENTS. SIGNAL AND BACKGROUND.

The measurements of the background (OFF) and measurements with reactor in operation mode (ON) were carried out for each detector position within single measuring period. The scheme of reactor operation and detector movements is shown in FIG. 29. A reactor cycle is 8-10 days long. Reactor shutdowns are 2-5 days long and usually alternate (2-5-2-...). The reactor shutdowns in summer for a long period for scheduled preventive maintenance. The movement of the detector to the next measuring position takes place in the middle of reactor operational cycle. Then the measurements are carried out at the same position until the middle of the next cycle. The scheme of detector operation and detector movements is shown on FIG. 29.

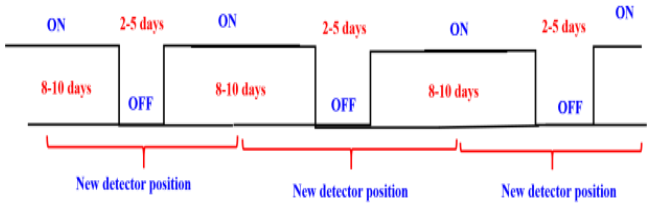

FIG. 29. Scheme of detector operation and detector movements.

The registration rate in the OFF regime of the reactor is mainly determined by the cosmic background and the way it changes with distance mainly depends on the structure of the building. At the distances 10-12 meter the blocking of cosmic rays is better because of concrete ceiling as shown in FIG. 2

The cosmic background is larger than the signal and it is the main problem of the experiment. It includes the correlated signal, the same the neutrinos produce. Therefore, distribution of cosmic background oscillations with time has to be accurately investigated.

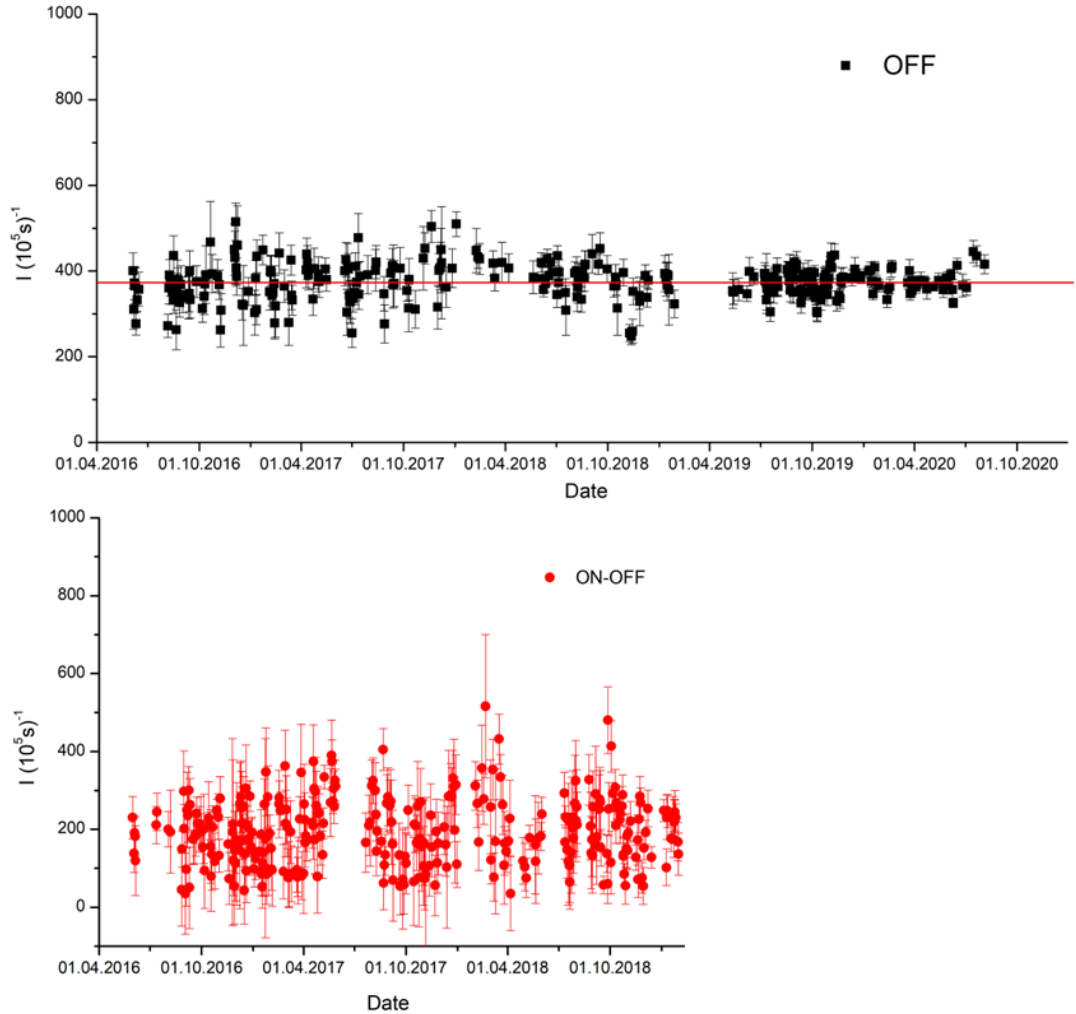

FIG. 30 The correlated signals produced by the cosmic background measured over the whole time (top). The correlated ON-OFF signals over the whole time (bottom). First cycle of measurements is from the beginning to the May 10, 2018. Second cycle is from May 19, 2018 to reactor stopping for the long period for reconstruction. Third cycle is measurements of the background during reactor reconstruction.

FIG. 30 (top) illustrates the pattern of the correlated background obtained during the whole measurement time. That pattern includes corrections for background changes caused by detector movements. The fluctuations of this pattern are mostly statistical. The measurements of the background were carried out by different time intervals, e.g. in 2020 when the reactor was not operating the measurement intervals were several times longer and hence the fluctuations are significantly lower. To perform a more detailed analysis one has to construct the distribution of the fluctuations normalized for each point at the corresponding statistical error. This distribution is shown in FIG. 31 (left). The standard deviation of this distribution is $1.07 \pm 0.01$, i.e. the additional fluctuations of cosmic background are only 7%.

The similar analysis of the stability for the ON-OFF signal is shown in FIG. 30 (bottom). The correction for the dependence of the ON-OFF signal on the distance is also applied. Distribution of the fluctuation of it is shown in FIG. 31 (right). As before, the fluctuations are normalized on the corresponding statistical uncertainty of the measurement. By the way it should be noted that reactor power fluctuations in different cycles is about 2% and also is averaged out in the long-term measurements. The standard deviation of this distribution is $1.09 \pm 0.02$, i.e. the additional fluctuations of cosmic background reactor power are only 9%.

The broadening in the statistical distribution of the neutrino signal due to fluctuations in the cosmic background and reactor power variation is 9% of the standard statistical deviation. As a result, we conclude that the measurements can be carried out with almost statistical accuracy despite the rather high cosmic background. In principle, one can add the correction on fluctuations of the cosmic background in each measurement using the fluctuations of atmospheric pressure. However, that is excessive since the correction is small and the fluctuations of the cosmic background averages out for the order of magnitude because the number of measurements is 87.

In conclusion, the analysis of the experimental data can be performed using the observed accuracy which is closed to statistical one.

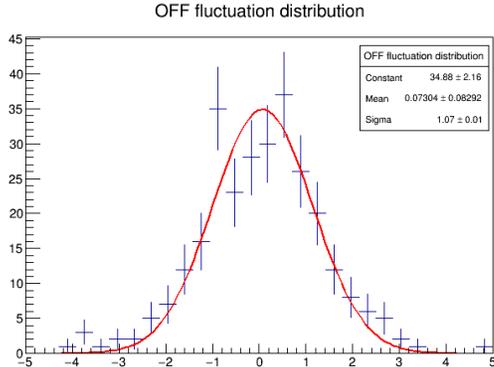
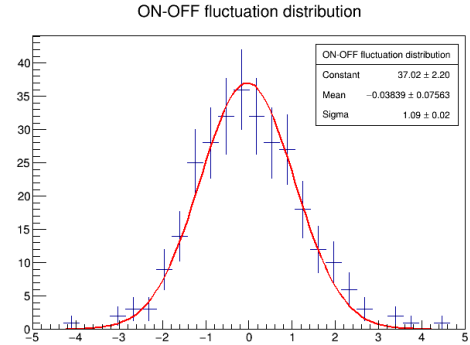

FIG. 31 The distribution of deviations from average value of correlated events rates background (OFF) and differences (ON-OFF) normalized on their statistical uncertainties.

## XIII.  SELECTION OF CORRELATED EVENTS

In order to form the reactor antineutrino signal one have to select correlated events. That selection technique including time intervals and topological criteria for IBD events are presented below. Every signal in 100 µs after an active shielding event is banned. Prompt (positron) and corresponding delayed (gamma from the neutron captured by gadolinium) signals have to occur in 300 µs time window. Events with time gap less than 100 µs between prompt and single delayed signal are considered as correlated. Events with time gap from 100 to 300 µs become a basis for estimation of accidental coincidence background which is subtracted for every measurement.

Prompt signal should be registered in single detector cell or in two adjacent cells. The threshold for prompt signal is 1500 keV, and for the last case energy t is applied to sum of two signals. Threshold for each signal is 200 keV. Delayed signal from few gamma-quanta of Gd(n, γ) reaction should be from 2 or more cells, which are not 3 cells far from prompt signal cell in any direction. That natural condition helps us to decrease accidental coincidence background. The experimental distribution of delayed signals after IBD event in section (4, 3) is shown in FIG. 32.

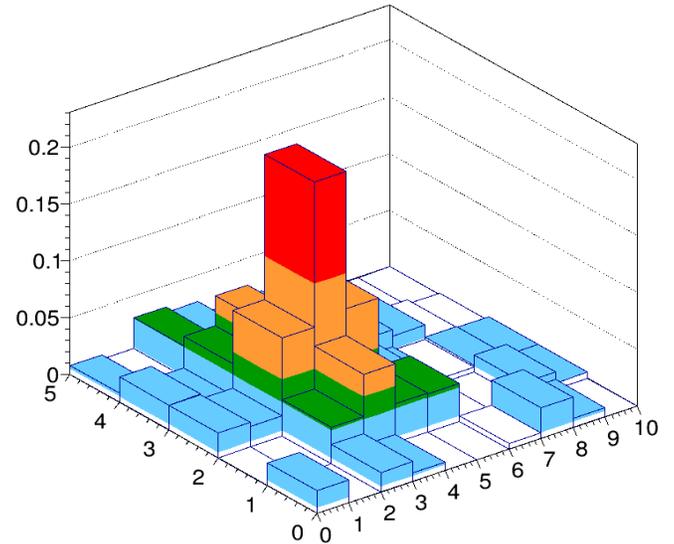

FIG. 32. Experimental distribution of delayed signals after the prompt signal occurred in section (4, 3).

## XIV.  BACKGROUND SPECTRUM

The background is one of the main problems of the experiment and it should be investigated in detail. The correlated background occurs due to fast neutrons according to the scheme shown in section VIII FIG. 13, where the prompt signal is formed by the recoil proton emitted in the elastic scattering of a neutron on hydrogen which is one of the main components of the scintillator. But the neutrons can also interact with carbon nuclei which also present in the scintillator. Moreover, neutrons can be inelastic scattered on the oxygen nuclei in the acrylic glass which is the main component of the mirror walls of the sections, and on the aluminum nuclei in the detector shell or section separations which are made of the aluminum alloy (see FIG. 19).

The spectrum of the background averaged over all the distances is shown in FIG. 33. The area 3-6 MeV has a bump and irregular behavior of the background dependences on the energy. (FIG. 33). That behavior can be explained by the structure of energy levels of the carbon, oxygen and aluminum nuclei. That irregularities of the spectrum superpose with the smooth spectrum function of the recoil protons emitted in elastic scattering of fast neutrons on the hydrogen and correlated background produced by unstable isotopes $^9$Li and $^8$He, which are the result of the interaction of cosmic rays (muons) with carbon nuclei. These isotopes decay in electron channel with following decay in which a neutron is emitted.

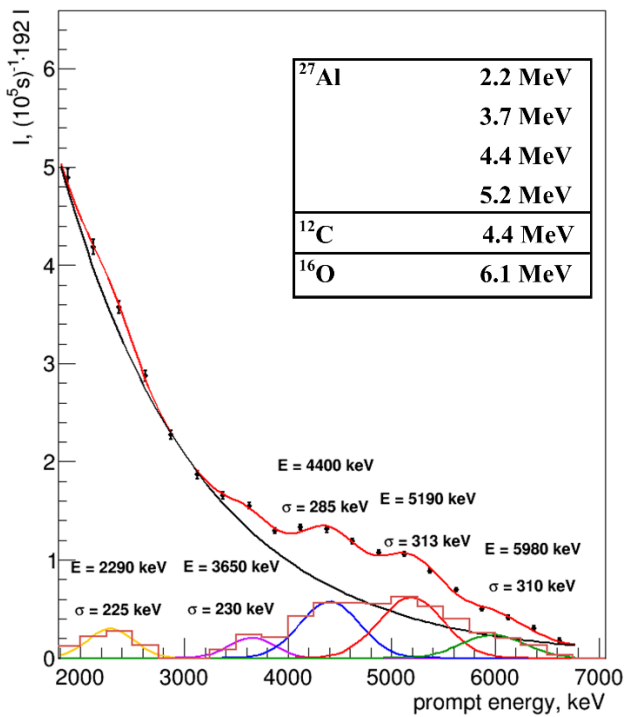

FIG. 33. Background spectrum averaged over all distances. Up - range 1500 ÷ 7000 keV, down- range 2500 ÷ 7000 keV.

A fast neutron in $(n, n')$ process leave the nucleus exited and its de-excitation occurs before the neutron can be thermalized and captured. The energies of the two lowest levels of the carbon nucleus are 4.44MeV and 7.65 MeV. The excitation of the level 7.65 MeV is possible with fast neutrons with energies above 8.25MeV. The level 7.66 is mainly decay in three alpha particles through the intermediate nucleus of $^8$Be [34]. The total energy of alphas is about 400 keV. The first exited state 4.44 MeV is a result of the inelastic scattering of the fast neutron and decay with emission of a gamma before the neutron can be thermalized and captured. Therefore, we can observe the correlated event. In case, of the oxygen nucleus the most intensive line has energy 6.13 MeV. The same scheme in case of the aluminum nucleus leads to gamma quanta with energies 2.2 MeV, 3.68 MeV, 4.4 MeV and 5.15 MeV.

At varying degree of confidence, it can be said that we observe the gammas with energies: 2.2 MeV, 3.68 MeV, 4.44 MeV, 5.15 MeV and 6.13 MeV.

Presence of that structure in the energy spectrum indicates that energy calibration of the detector was the same in all measurements. Moreover, since we observe the separate lines in the spectrum, however the resolution is rather low, we can estimate the energy resolution of the detector and suggest the decomposition of the background of the background spectrum into the set of gamma lines (see FIG. 33). In range from 2.0 MeV to 7.0 MeV the energy resolution changes from 225 keV to 310 keV. The estimated value corresponds to the resolution of the whole detector, which is better than the resolution of the single section, because of the taking into account the signals in adjacent section.

In order to verify the estimation of the detector energy resolution that we used in the decomposition of the measured background energy spectrum we performed MC calculation of the energy resolution of the whole detector. The examples of the simulation results for gammas with energies 2.3 MeV, 4.4 MeV and 6 MeV are shown in FIG. 26. The calculated resolution is $2\sigma=250$ keV at energy 4.4 MeV. But in the experiment the resolution at this energy is $2\sigma=570$ keV, i.e. two times worse. The partitions between the sections, difference in calibration of the sections and instability in the three years long measurements can decrease the energy resolution.

## XV. ANTINEUTRINO SPECTRUM (ON – OFF) AND SPECTRUM OF ACCIDENTAL COINCIDENT

In order to obtain antineutrino spectrum as the difference of ON-OFF spectra we subtract contributions of both processes with uncertainty equal to fluctuations of cosmic background, which, as was discussed before, are only 7% of the statistical accuracy of a single measurement and they are averaged out for an order of magnitude during the long term measurements. The ON-OFF difference is 223 events per day in distance range 6 – 9 m. Signal/background ratio is 0.52.

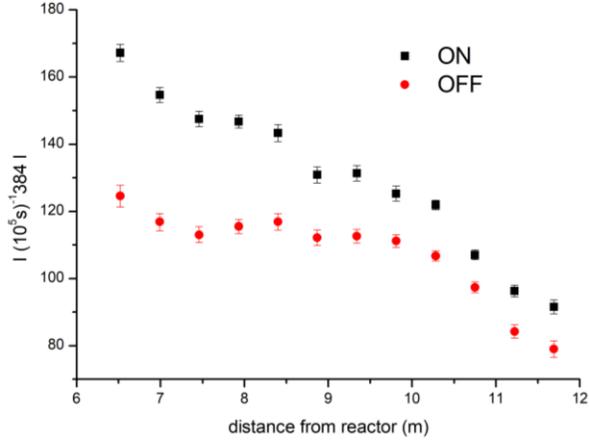

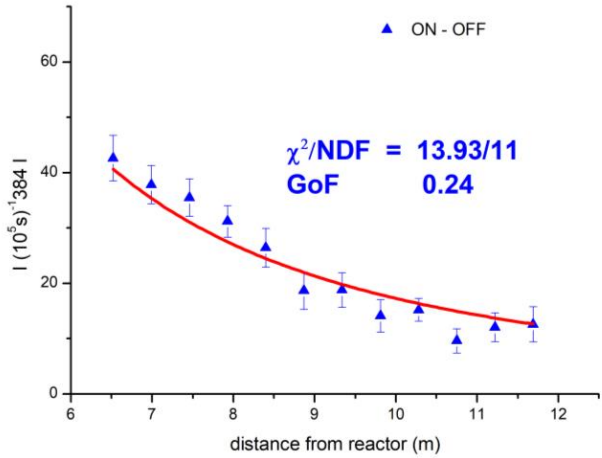

FIG. 34. Up – event rate with operating (ON) and stopped (OFF) reactor. Down – difference between event rates (ON-OFF) as functions of distance.

FIG. 34 illustrates two spectra of correlated signals with reactor ON and OFF (background), and also the ON-OFF spectrum. As was discussed above their difference describes antineutrino spectrum of the reactor. The accuracy of that statement is based on the fact that changes of fast neutron flux on the surface of containment building of the reactor which can contribute in ON-OFF difference do not exceed $(1.1 \pm 0.45)\%$ as discussed in section V. Moreover, this is not a problem for these measurements, especially since this background cannot have the oscillation behavior. Event rate with operating (ON) and stopped (OFF) reactor, and also the difference of that rates (ON-OFF) as functions of distance are shown in FIG. 34.

FIG. 35 (top) illustrates the ON and OFF spectra averaged over all distances. FIG. 35 (bottom) illustrates ON-OFF signals averaged over all distances – blue histogram and the difference of averaged spectra for ON and OFF signals – red histogram. Notice, that in the analysis of antineutrinoi signal one have to use ON-OFF signal for each distance obtained in the nearest measurements. That way one obtains the best possible compensation of the fluctuations of the cosmic background with time.

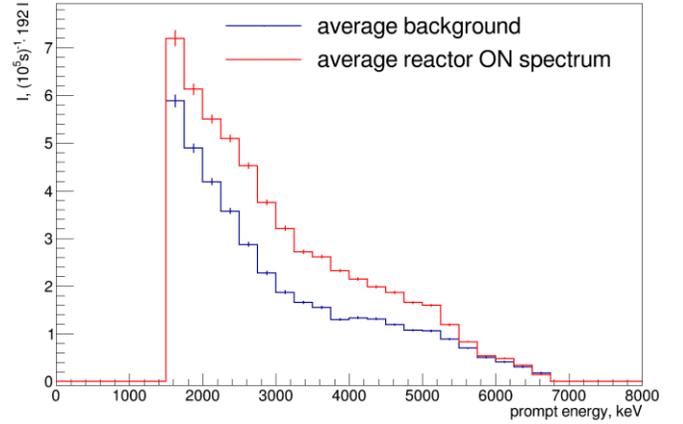

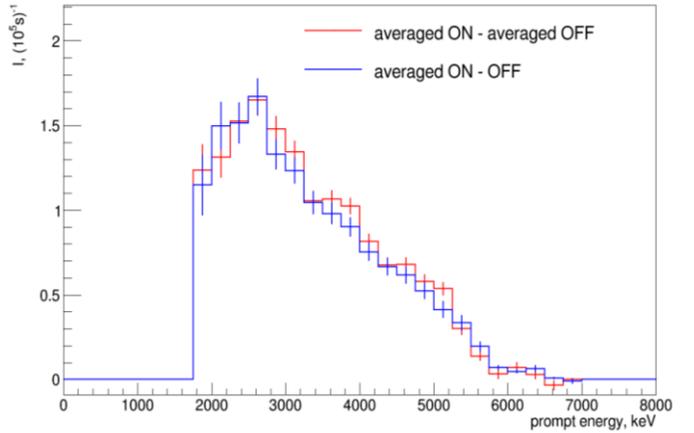

FIG. 35. Top - the ON and OFF spectra averaged over all distances. Bottom - ON-OFF signals averaged over all distances – blue histogram and the difference of averaged spectra for ON and OFF signals – red histogram.

Besides the correlated background there is also a problem of accidental coincidence background. The energy spectrum of the accidental coincidence background is shown in FIG. 36 (up) for three distances. The influence of reactor operation mode on the accidental coincidence background is also shown in FIG. 36 (down). The background significantly increases if we decrease the limit of delayed signals under 3 MeV level and even some dependence of reactor mode can be observed. As was discussed before, this problem can be solved by setting lower limit of delayed signal to be 3.2 MeV, because neutron capture by Gd yields signal with sufficiently high energy up to 8MeV, while natural radioactivity background is almost zero above 3 MeV. The limit of prompt signal we set at level 1.5 MeV to significantly decrease the amount of cut off of neutrino events.

Thus, conclusion of this analysis is selection of the optimal lower thresholds for prompt and delayed signals: 1.5 MeV and 3.2 MeV respectively.

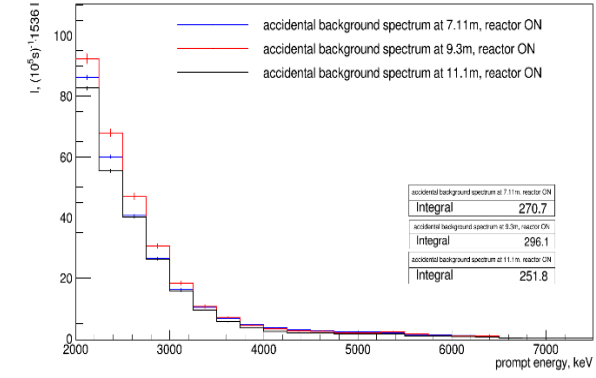

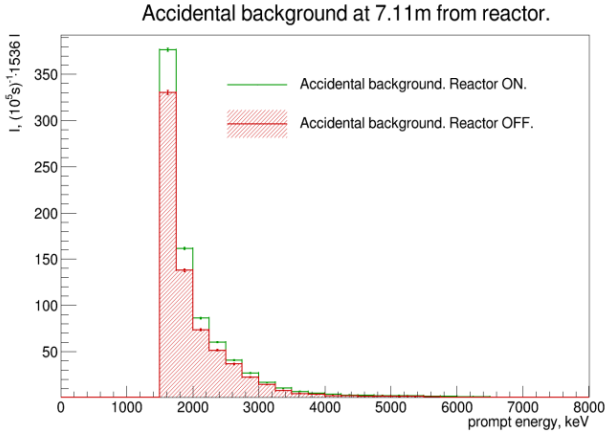

FIG. 36. The spectra of accidental coincidence background at 7.11m, 9.3m, and 11.1m (up). The ON and OFF spectra at the point closest to the reactor - 7.11m (down).

## XVI. COMPARISON OF EXPERIMENTAL ANTINEUTRINO SPECTRUM WITH CALCULATED REACTOR SPECTRUM

In order to compare the experimental spectrum of antineutrinos with the calculated spectrum of reactor antineutrinos one has to consider the results of MC calculations of efficiency of registration of IBD process in the detector. The spectrum obtained in the experiment corrected by the efficiency factor should be compared with expected spectrum of prompt signals calculated in the MC simulation. An example of such comparison is shown in FIG. 37, where we present the experimental spectrum of prompt signals, averaged over all distances for better statistical accuracy, and the MC spectrum of prompt signals, obtained using the spectrum of $^{235}$U and taking into consideration the threshold of experimental signals.

As was already pointed out, the correct way of analysis for neutrino oscillations implies usage of the averaged ON-OFF signal spectrum for each distance, moreover, it had better be obtained in the nearest measurements, in order to better compensate for temporal fluctuations of the cosmic background. However, a comparison can also be made using the difference between the ON and OFF spectra averaged over all distances. This allows to include in the data processing long-term background measurements obtained in 2020. Both spectra then can be compared with the calculated one. Results of such comparison are shown in FIG. 37. Compensation of fluctuations of the cosmic background is better for the measurements with short time between reactor ON and reactor OFF period.

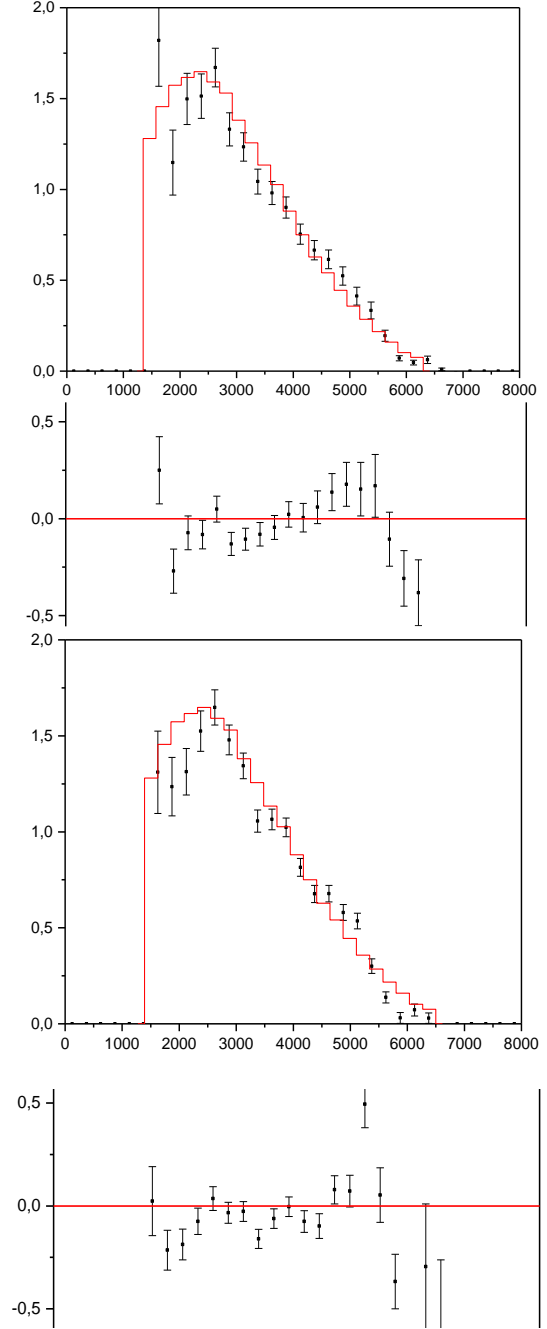

FIG. 37 Comparison of the calculated MC spectrum of antineutrino for $^{235}$U with the experimental ON-OFF spectrum. Below their difference normalized to the calculated spectrum is shown. The first two figures represent the ON-OFF signal spectrum averaged over all distances, the second two figures represent the difference between the ON and OFF spectra averaged over all distances, where long-term background measurements are included.

The comparison of the experimental and calculated spectra demonstrates slight deviation from the theoretical picture within current statistical uncertainty. The reasons for the possible discrepancy may be due to different circumstances. First, the energy calibration of the calculated neutrino spectrum is not fixed, and the energy dependence of the detection efficiency is not completely accurate. Second, one can consider the presence in the spectrum of the often discussed bump in the 5MeV region. Both reasons cannot be ruled out. To avoid this problem, a spectrally independent method for analyzing experimental data is needed. Further we propose such analysis method.

## XVII. SPECTRAL INDEPENDENT METHOD FOR ANALYSIS OF EXPERIMEHTAL DATA

Here we suggest spectral independent method of data analysis based on relative measurements, in which the spectrum is cancelled out. Therefore, it is a model independent analysis of data. It is based on equation (2), where in the left hand side the numerator is the rate of antineutrino events in certain energy area at certain distance with correction to geometric factor $L^2$ and denominator is the rate of antineutrino events with the same energy, but it is averaged over all distances:

$$R_{ik}^{\exp} = (N_{ik} \pm \Delta N_{ik})L_k^2 \Big/ K^{-1} \sum_k^K (N_{ik} \pm \Delta N_{ik})L_k^2 =$$

$$= \frac{\left\langle S(E)\left(1 - \sin^2 2\theta_{14} \sin^2\left(\frac{1.27\Delta m_{14}^2 L_k}{E}\right)\right)\right\rangle_i}{K^{-1}\sum_k^K \left\langle S(E)\left(1 - \sin^2 2\theta_{14} \sin^2\left(\frac{1.27\Delta m_{14}^2 L_k}{E}\right)\right)\right\rangle_i}$$

$S(E)$ – initial $^{235}$U spectrum, $\langle s \rangle$ – the integration with function of energy resolution with $\sigma = 250$ keV and integration over energy bin intervals. The right part of the equation is the same ratio written in analytical form taking into account the oscillation hypothesis. There was estimation that showed the right hand side of the equation practically does not depend on function $S(E)$ [28].

Denominator of the ratio is the rate of antineutrino events with the same energy, but it is averaged over all distances, hence oscillation effect is considerably averaged out in denominator if oscillations are frequent enough in considered distances range. In this case it is:

$$R_{ik}^{\text{th}} \approx \frac{1 - \sin^2 2\theta_{14} \sin^2(1.27\Delta m_{14}^2 L_k/E_i)}{1 - \frac{1}{2}\sin^2 2\theta_{14}} \xrightarrow[\theta_{14}=0]{} 1, \quad (3).$$

In no oscillation case this expression becomes equal to unity. $R_{ik}^{\exp}$ ratio in approximation (3) with accuracy to coefficient $1/(1 - 1/2\sin^2 2\theta_{14})$ corresponds to the equation (1) which is describing the process of oscillations.

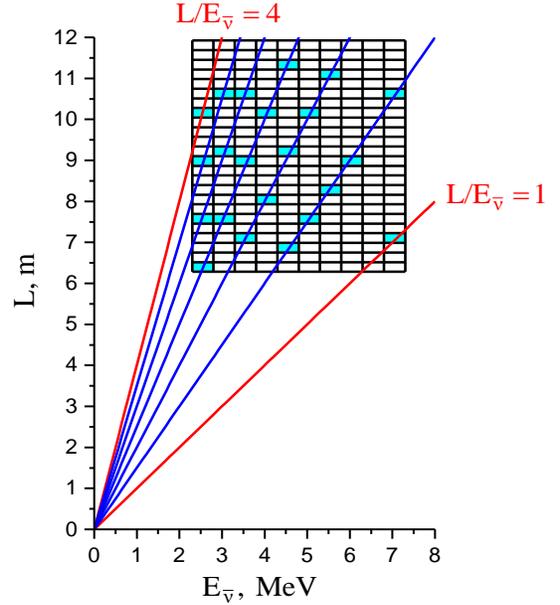

FIG. 38 Illustration of the method of coherent summation of measurement results to obtain the dependence of $R_{ik}^{\text{sim}}$ on the ratio L/E.

As will be shown later, the oscillation parameters found in the experiment satisfy approximation (3), although equation (2) is used in all procedures of the oscillation analysis.

At the end of this section, it should be emphasized that the proposed data processing method greatly simplifies the requirements for the energy calibration of the detector and determining its efficiency, as follows from the $R_{ik}^{\exp}$ ratio. Moreover, this method allows to directly demonstrate the process of oscillations.

The moveable detector method is also a method of relative measurements, so their joint use significantly increases the reliability in revelation of the oscillation process.

FIG. 38 demonstrates the scheme which is applied to construct the dependence on the parameter L/E using the matrix of the experimental data and the process of coherent summation

## XVIII. MONTE CARLO SIMULATION OF RESULTS EXPECTED WITH EMPLOYING OF SPECTRAL INDEPENDENT METHOD OF DATA ANALYSIS

In this section we present results of the MC simulation in which we incorporated geometric configuration of the antineutrino source and detector including the sectioning. For example, in this simulation we have used parameters $\Delta m_{14}^2$ and $\sin^2 2\theta_{14}$, close to the values that will be derived from the analysis of experimental data later. The goal of this simulation is to see what the process of oscillations on the plane $(E, L)$ looks like and how to extract the process of oscillations as a dependence on the L/E ratio according to the equation (1).

The source of antineutrino with geometrical dimensions of the reactor core 42x42x35cm³ was simulated, as well as a detector of antineutrino taking into account its geometrical dimensions (50 sections of 22.5x22.5x75cm³). The antineutrino spectrum of $U^{235}$ (though it does not matter since energy spectrum in equation (2) is cancelled out) factored by function of oscillations $1 - \sin^2 2\theta_{14} \sin^2(1.27\Delta m_{14}^2 L_k/E_i)$ was used.

The expected oscillation effect for the different energy resolution of detector is shown in FIG. 39 and FIG. 40 on the top as the plane $(E, L)$, and at the bottom as a function of the L/E ratio obtained by adding data with the same L/E ratio.

This procedure was illustrated in FIG. 38. FIG. 39 and FIG. 40 (on the top) illustrate the simulated matrix of $R_{ik}^{sim}$ ratio $(N_{ik}L_k^2/K^{-1}\sum N_{ik}L_k^2)$, where the oscillation process according to the formula (1) is considered. In simulation the statistical accuracy of ratio $\Delta N_{ik}/N_{ik}$ equal to 1%, which is significantly better than the experimental value.

The most important parameter in this simulation was the energy resolution of the detector, which is determined by the energy step in the matrix 250 keV and 500 keV correspondingly. It can be seen that the degradation of the energy resolution of the detector suppresses the observed effect of oscillations, but the number of observed oscillation periods decreases, and the amplitude of the first observed oscillations is still the same.

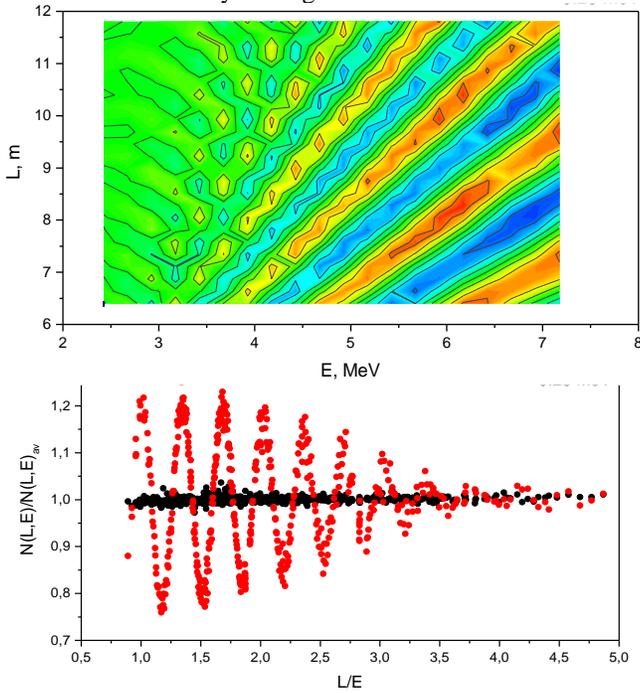

FIG. 39 The simulated matrix of $R_{ik}^{sim}$ − ratio $(N_{ik}L_k^2/K^{-1}\sum N_{ik}L_k^2)$ for energy resolutions of detector ±125keV (top); dependence of $R_{ik}^{sim}$ from L/E ratio (bottom).

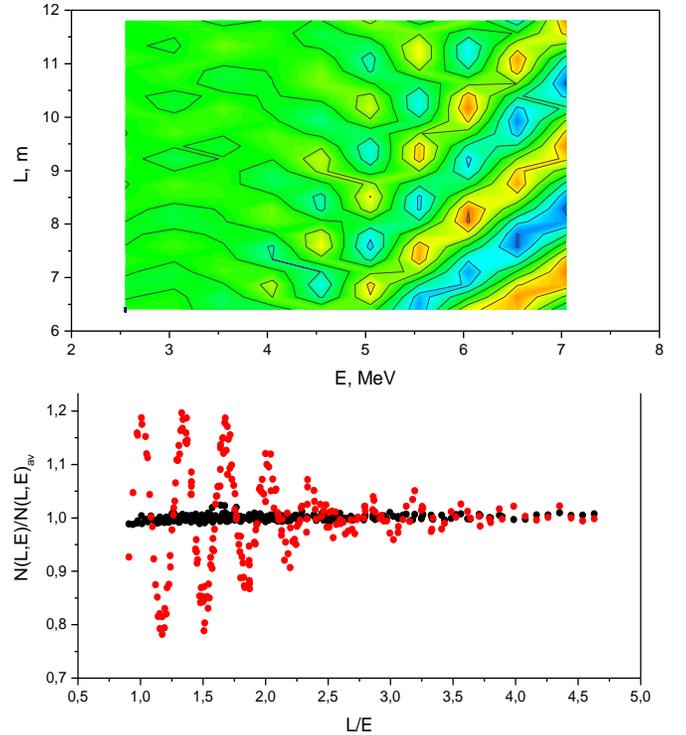

FIG. 40. The simulated matrix of $R_{ik}^{sim}$ − ratio $(N_{ik}L_k^2/K^{-1}\sum N_{ik}L_k^2)$ for energy resolutions of detector ±250 keV (top); dependence of $R_{ik}^{sim}$ from L/E ratio (bottom).

The MC simulation results can be summarized in several conclusions. First of all, it is possible to get an idea of that as the effect of oscillations for $R_{ik}^{exp}$ relation on the $(E, L)$ plane looks. Secondly it becomes clear how to directly observe the effect of oscillations.

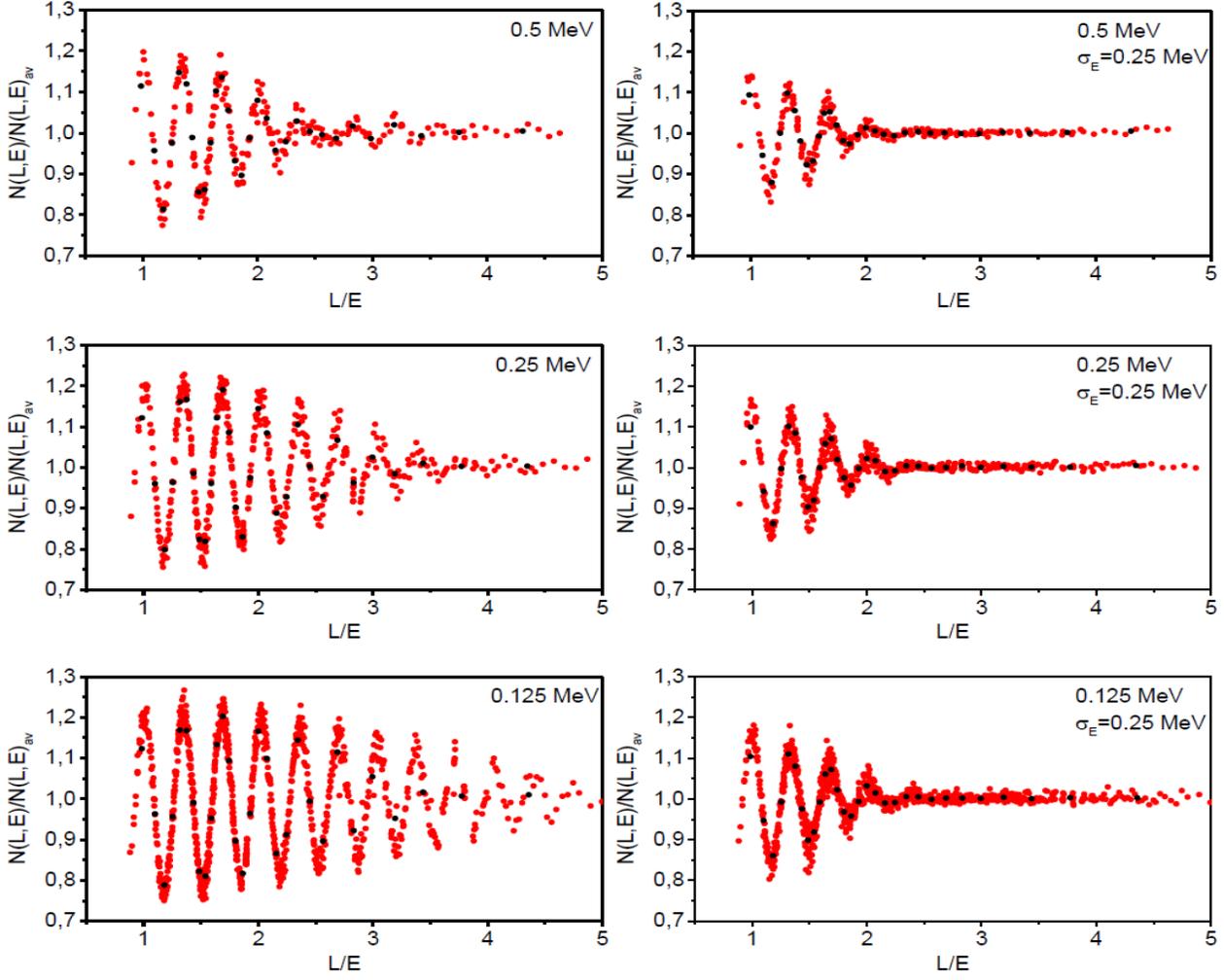

FIG. 41. On the left - the simulation of oscillation curve in assumption of the ideal energy resolution and intervals 125 keV, 250 keV, and 500 keV. On the right – the simulation of oscillation curves with energy resolution of the detector with $\sigma = 250$ keV and intervals 125, 250, 500 keV.

In this section we would like to clarify some aspects of experimental data processing and the MC simulations. The problem of energy resolution requires special attention. The construction of the matrix of measurements on the $(E, L)$ plane with energy interval $\Delta E$ includes the energy resolution. The only difference between results presented in the FIG. 39 and FIG. 40 is the energy bin intervals which were $\Delta E = 250$ keV and $\Delta E = 500$ keV. The original simulation of the data set was performed with assumption of the ideal detector energy resolution. In order to correctly choose the energy interval we performed the simulation described below.

We simulated the experiment with energy resolution σ=±250 keV. The data processing is performed with energy intervals 125 keV, 250 keV and 500 keV. For comparison we also performed calculations with ideal energy resolution and different energy intervals.

The comparison of left and right images in FIG. 41 demonstrates the influence of detector energy resolution on the oscillation curve. The right images shows that if one set the energy interval smaller or equal to energy resolution σ=250 keV then that choice have almost no effect on the oscillation curve. If one set the interval to be $\Delta E$=500 keV then it substitutes the detector energy resolution σ=250 keV and such interval can be used for fast analysis of the data. In the final analysis one have to use the actual detector energy resolution σ=250 keV. The simultaneous usage of energy interval $\Delta E$=500 keV and energy resolution σ=250 keV is incorrect, because it includes into the analysis the resolution of the detector twice as it was done in the work [35].

In summary, on the basis that estimated detector energy resolution for the gamma quanta registration is in the range 220 – 310 keV (see FIG. 33) and features of the positron registration (see FIG. 25) we argue that the right way to process the data for the oscillation analysis is by usage the energy resolution of the detector σ=250 keV which does not depend on the energy of a positron.

## XIX. THE MATRIX OF MEASUREMENTS

The results of experimental measurements of the antineurino flux dependence on distance and energy of antineurino can be presented in the form of a matrix, where $N_{ik}$ is difference of ON - OFF rates for the i-th interval of energy and for the k-th distance from the reactor core. In total there are 24 positions of antineurino spectral flux measurements from 6.4 m to 11.9 m. The distance step corresponds to the cell size of 23cm, which is twice smaller than reactor core size.

The energy spectrum is divided using various intervals such as 500 keV, 250 keV, 125 keV. The choice of a smaller interval is unreasonable because it would lead to decrease of statistical accuracy of calculated matrix elements. Even with 500 keV interval the obtained statistical accuracy is insufficient to construct the oscillation curve. The large statistical uncertainties obscure the the variations in the oscillation amplitude. Therefore, we have to sum the close points of the L/E dependence, like it was shown schematically in FIG. 38. This procedure is valid if the summation do not suppress the amplitude of the oscillation.

In further analysis we use the 500 keV interval and sum the groups of 8 adjacent points of the L/E dependence. In addiition, we use 250 keV intervals and 16 point groups, and finally 125 keV intervals and 32 point groups. The black points at the curve (FIG. 41) are the result of summation of the points.

This exact case is illustrated in the right image in FIG. 41 (right) The number of red points corresponds to the number of matrix elements. The black points are the result of the summation of close points of the L/E dependence. The number of black points is the same for all three curves. These three ways of data processing use almost the same data and the results are equal within statistical accuracy. The processing of the same data three times is reasonable because it helps to avoid the problem of fluctuations in the random data sample by averaging over the results of three described analysises. It is important that the statistical accuracy of the averaged result do remains the same. The total results of the described data analysis procedure are listed below.

For comparison with theory, the MC simulation of the experiment should be carried out and the energy resolution of the detector should be taken into account. As mentioned in the previous section, the processing of data for the analysis of the oscillation process should be carried out using the energy resolution of the detector σ = 250 keV, independent of the positron energy.

## XX. MEASUREMENTS

Here we present the results of the analysis of all data collected from June 2016 till October 2020. On July 2019 reactor was stopped for renovation. However, background has been measured from July 2019 till October 2020. It was of great importance since reactor shutdowns are 2-3 times shorter than the operational periods. Measurements from September 2018 to July 2019 were carried out mainly in near positions to the reactor, where the signal to background ratio is significantly better. Overall measurements with the reactor ON were carried out for 720 days, and with the reactor OFF- for 860 days. Reactor on and off operations occurred 87 times.

It is convenient to divide the measurements in three measurement cycles. The first cycle was carried out between June 2016 and May 2018. The results were published in ref. [27]. It was the first observation of the oscillation effect at the level of three standard deviations. The second cycle was carried out from May 2018 to October 2019. During that period the statistical accuracy was increased by the factor 1.4. The result of processing of the data obtained in both periods confirmed the oscillation effect with the same confidence level. The results were published in ref. [36]. In the third cycle only the measurements of the background were carried out. In this 1.5 year long period the reactor was stopped for reconstruction. The data obtained in the third measurement cycle are currently included in this article.

Below we present the results of the described above scheme of the analysis applied to the data obtained in the first measurement cycle, the data obtained in the first and second cycles, and the data obtained in all three measurement cycles.

It is necessary to notice that measurements in the mode when reactor ON and reactor OFF periods are close in time are preferable. In this case fluctuations of the background are compensated. If we use long and independent measurements of the background and we compare them to measurements obtained with operating reactor with an interval of 1-2 years, then we can observe the fluctuations of the background data considerably outside statistical accuracy. As it will be shown further, usage of long term measurements during the reactor reconstruction is unreasonable if we aim to observe the oscillation effect. It leads only to increase, but not reduction of the uncertainties of measurements. Nevertheless, these measurements were very useful to detailed study of the background.

## XXI. THE FIRST PHASE OF DATA ANALYSIS

Here we present the data analysis of reactor antineutrino signal for the first and second measurement cycles combined.

Initial distribution of the count rate ON- OFF = $N_\nu$ in the entire energy range is shown in FIG. 42. It is count rate deviation from the mean value for the different series of measurements normalized to its statistical uncertainties. It allows combine all measurement results at different distances to find additional dispersion beside statistical. As can be seen from FIG. 42 (left), it shows a normal distribution determined practically by statistics. It means that additional sources of instabilities besides the cosmic background fluctuations are absent.

We compare it with the distribution obtained for the ratio $R_{ik}^{\exp}$ for the same dataset. It is difference of $R_{ik}^{\exp}$ from 1, and as well as the distribution ON-OFF, normalized by $\sigma$.

In this case whole energy range was divided into 9 intervals (500 keV) and distances were divided into 24 intervals, so matrix has 216 elements. FIG. 42 (right) shows the distribution of all 216 points over the L/E range from 0.9 to 4.7. One can see that the distribution $R_{ik}^{exp}$ already differs from normal. The value of the $\chi^2$/dof parameter is 25.9/16 which disfavors this function because confidence level for this result is only 5%. Additional width of $R_{ik}^{exp}$ distribution could appear due to oscillation effect. It can be considered as first evidence of oscillation effect, which broadens this distribution.

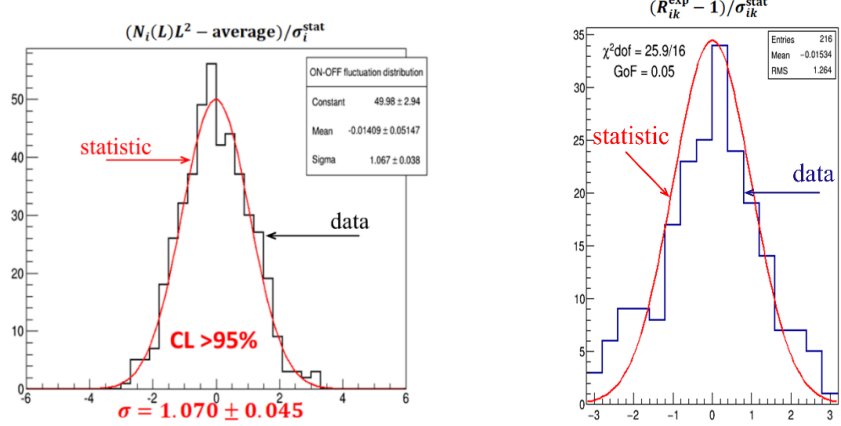

FIG. 42. Left - distribution of the count rate ON- OFF in the entire energy range, normalized by $\sigma$. Right - distribution $R_{ik}^{exp}$ of all 216 points over the L/E range from 0.9 to 4.7, normalized by $\sigma$.

## XXII. THE ANALYSIS OF OSCILLATION PROCESS ON THE PLANE $\Delta m_{14}^2$, $\sin^2 2\theta_{14}$

The most optimal oscillation parameters can be found by using the expression:

$\sum_{i,k}\left(R_{ik}^{exp} - R_{ik}^{th}\right)^2 / \left(\Delta R_{ik}^{exp}\right)^2 = \chi^2(\sin^2 2\theta_{14}, \Delta m_{14}^2)$.

The matrix of measuremets should be compared with the MC calculated matrix for detector energy resolution ±250 keV. Below we present the search for oscillation parameters on the plane $\Delta m_{14}^2$, $\sin^2 2\theta_{14}$ using $\Delta\chi^2$ method and the dependence of the R ratio on the parameter L/E, i.e. the oscillation curve.

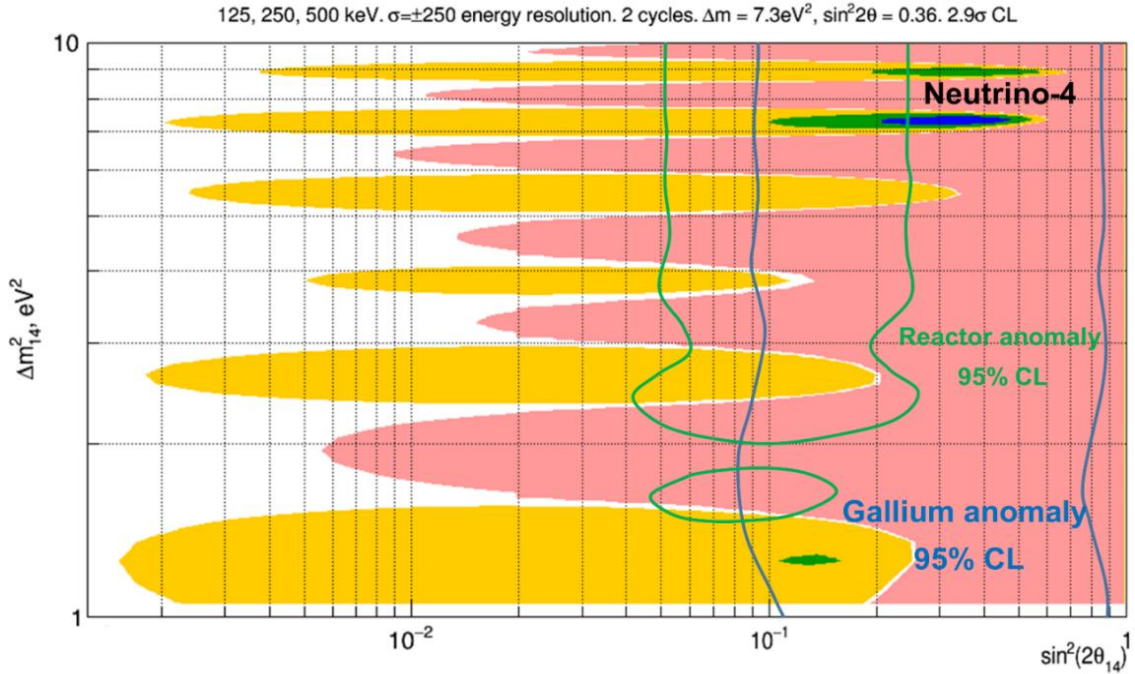

FIG. 43. The results of the analysis of data obtained in first and second measurement cycles on the plane $\Delta m_{14}^2$, $\sin^2 2\theta_{14}$.

The results of the analysis of data obtained in first and second measurement cycles on the plane $\Delta m_{14}^2, \sin^2 2\theta_{14}$ are presented in FIG. 43. The colored in pink area of oscillation parameters is excluded with CL more than 3σ. However, in the area $\Delta m_{14}^2 = 7.3 eV^2$ and $\sin^2 2\theta_{14} = 0.36 \pm 0.12_{stat}$ the oscillation effect is observed at 2.9 σ CL. Here we also illustrate the areas of the Reactor anomaly and the Gallium anomaly.

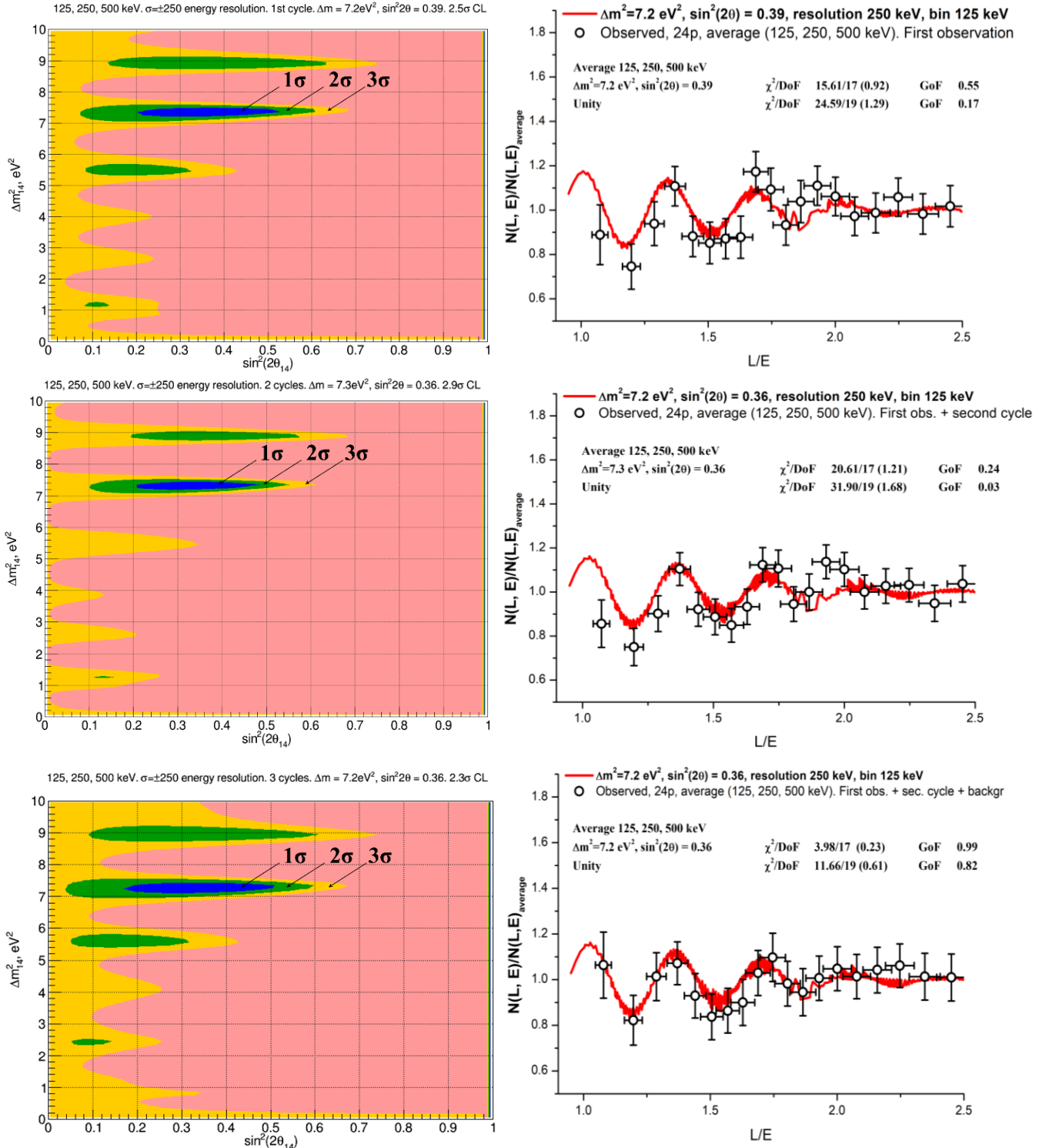

FIG. 44. Top – results of data analysis of the first measurement cycle (I), middle – the results of data analysis of the first measurement cycle together with the second measurement cycle (II), bottom – results of data analysis of the first measurement cycle, with the second measurement cycle and with the third measurement cycle (III). On the left is the central part of the region $\Delta m_{14}^2, \sin^2 2\theta_{14}$ and on the right is the oscillation curve.

The FIG. 44 shows the results of the data analysis of the first measurement cycle (I), the results of the data analysis of the first measurement cycle together with the second measurement cycle (II), and the results of the data analysis of the first measurement cycle, with the second measurement cycle and with the third measurement cycle (III). On the left is the part of the region on the $\Delta m_{14}^2, \sin^2 2\theta_{14}$ plane and on the right is the oscillation curve. It should be noted, that due to error increased goodness of fit criterium is closing to unity, which is natural if effect is exist, but increasing of errors is out of the statistical limitation.

In Table III we list the oscillation parameters, $\chi^2$/DoF and GoF for all three cases (I, II, and III);

TABLE III

| Case | Best fit oscillation parameters $\Delta m_{14}^2, \sin^2 2\theta_{14}$ | $\chi^2$/DoF (Reduced $\chi^2$) fit w/ and w/o oscillation | Goodness of fit w/ and w/o oscillation |
|---|---|---|---|
| I | 7.2 eV$^2$, 0.39 (2.5 σ) | **15.61/17 (0.92)** 24.59/19 (1.29) | 0.55 0.17 |
| II | 7.3 eV$^2$, 0.36 (2.9 σ) | **20.61/17 (1.21)** 31.90/19 (1.68) | 0.24 0.03 |
| III | 7.2 eV$^2$, 0.36 (2.3 σ) | **3.98/17 (0.23)** 11.66/19 (0.61) | 0.99 0.82 |

It is important to emphasize that the statistical uncertainty of the parameter $\sin^2 2\theta_{14}$ decreased after the inclusion of the data of the second measurement cycle, while the CL increased. However, if we include in the analysis additional measurements of the background obtained in the third cycle the CL decrease and the statistical uncertainty of the parameter $\sin^2 2\theta_{14}$ increase. It was mentioned above, that measurements with short intervals between signal and background measurements are preferable. In this case, the background fluctuations are compensated. If we compare the results of background measurements with the results obtained with the reactor in operation mode which have time interval of 1-2 years then we can get the deviation of the background which exceed the statistical accuracy. It leads only to increase, not a decrease, in measurement uncertainty. Therefore, we must focus on the results obtained in the first and second cycles of measurements. $\Delta m_{14}^2 = 7.3 \pm 0.13_{st} \pm 1.16_{sys} = 7.3 \pm 1.17, \sin^2 2\theta = 0.36 \pm 0.12 (2.9\sigma)$

## XXIII. ANALYSIS OF POSSIBLE SYSTEMATIC EFFECTS

1.The study of possible systematic effects was performed using a background of fast neutrons created by cosmic rays. In order to study systematic effects, one has to turn off antineutrino flux (turn off the reactor) and perform the same analysis of the collected data. That procedure has sufficient precision since we have collected significant amount of background data. Result of background measurements analysis for oscillation effect manifestation on the plane $\Delta m_{14}^2$ and $\sin^2 2\theta_{14}$ is shown in figure FIG. 45. The data of all measurement cycles are combined.

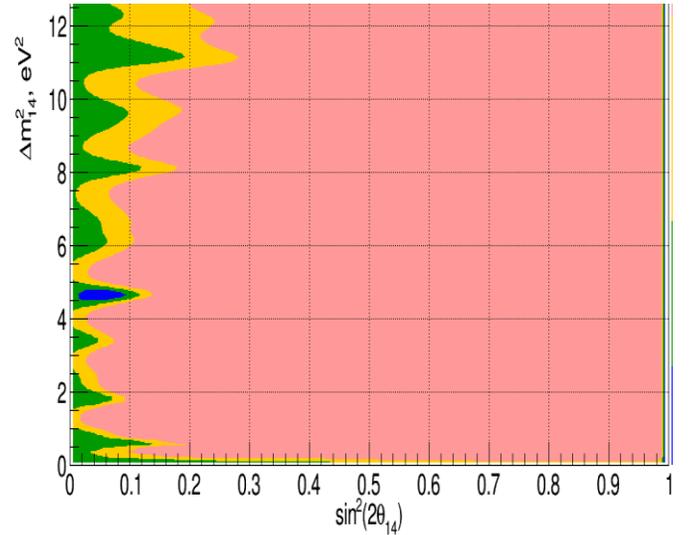

FIG. 45. Data analysis results for the background on the plane $\Delta m_{14}^2$ and $\sin^2 2\theta_{14}$.

The results of the background analysis presented in FIG. 46 (bottom), which show that the background cannot form the signal observed for ON-OFF data (FIG. 46 top).

Correlated background (fast neutrons from cosmic rays) slightly depends on distances from reactor due to inequality of concrete elements of the building. If, earlier during constructing the R-ratio for the neutrino signal, we introduced a correction for the distance $L^2$, now we used the dependence of the background on the distance, shown in FIG. 34.

From FIG. 46 it can be seen that the dependence of the R-ratio on L/E satisfies a constant fit with $\chi^2/DoF = 1.3$, when the oscillation fit gives $\chi^2/DoF = 6.1$. Therefore, we can conclude that the apparatus does not produce systematical effect.

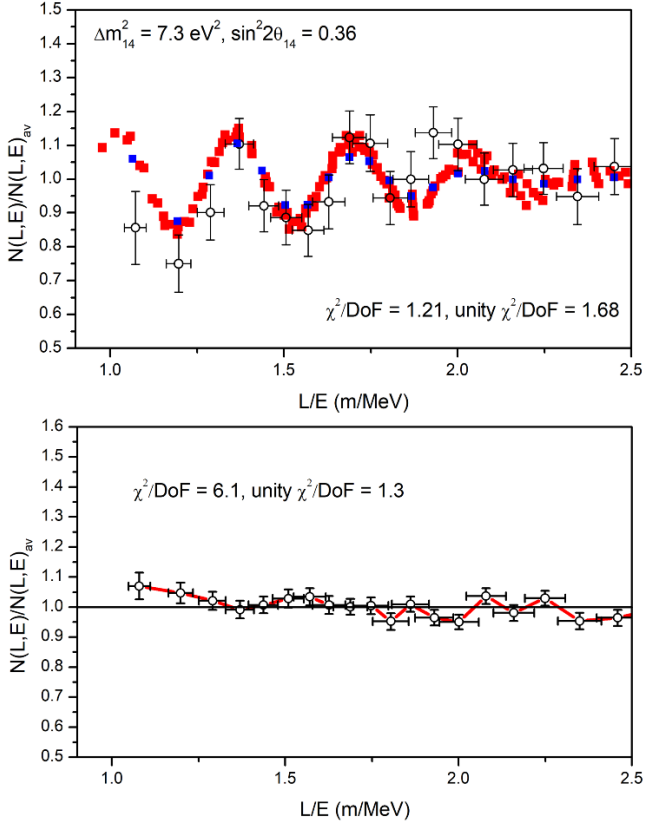

FIG. 46. Comparison of the R-ratio versus L/E for the neutrino signal (top) and the R-ratio versus L/E for the background (bottom).

It is shown above that it is unreasonable to add corrections into the oscillation curve. Also, the square root of the sum of squares of significant uncertainties of points at the oscillation curve and small uncertainties at the background curve is very close to the uncertainties at the oscillation curve (deviation is about 2-3% of the uncertainties at the oscillation curve).

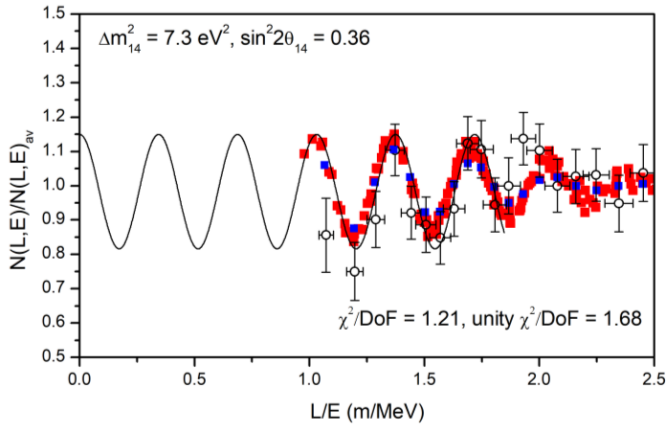

FIG. 47 Complete curve of the oscillation process starting from the reactor core center.

2. In addition, it should be noted that the experimental points should be fitted with such a sinusoidal dependence, which has a maximum at the origin, since the process of oscillations starts from the source. This significantly reduces the set of sinusoids available for fitting. FIG. 47 shows the complete curve of the oscillation process starting from the reactor. There is a maximum at zero, from here the process of oscillations begins.

3. Below we are trying to perform additional search for systematic effects connected with the correlated background.

We would like to remind that correlated background arise from fast neutrons as a result of elastic and inelastic scattering $n, n'$. In the reaction of elastic scattering on hydrogen, a recoil proton appears which imitates the signal from a positron. The reaction of inelastic scattering of fast neutrons occurs on carbon nuclei. The fast neutron excites the carbon nucleus, which de-excites before the neutron is thermalized and captured. This process produces a correlated event similar to the IBD process. This looks rather dangerous for the search for neutrino oscillations. However, it should be clarified that the background spectrum is subtracted when we form the ON-OFF difference signal to detect neutrino events. Again, one may be concerned that this subtraction may not be complete due to the effect of background fluctuations associated with fluctuations in atmospheric pressure and temperature since measurements with the reactor running and with the switched off states occur at different times. However, it is possible to quantify this incomplete compensation of the effect. In our case, the measurements have been carried out about for about 4 years and the atmospheric pressure variations that is $\pm 1.1\%$ were nicely averaged. Since the reactor on and off operations occurred 87 times, the average contribution of cosmic background fluctuations to the measurement results does not exceed $\pm 0.1\%$ or approximately $\pm(0.3 \div 0.5)\%$ with respect to the neutrino signal, the oscillations of which are $\pm(10 \div 15)\%$.

So, background fluctuations uncorrelated with the reactor power do not create a danger of a false oscillation effect. Therefore, it is necessary to investigate the possible change in any parameters, correlated with the reactor power. For example, the temperature in the laboratory increases when the reactor is operating. If in this case, for example, the gain of the PMT changes, this will lead to a shift in the spectra. Irregularities in the background spectrum will be shifted. When the background is subtracted, a difference with a periodic structure appears. This reasoning raises the following concern about the possibility of a false oscillation effect. This requires a quantitative assessment of the possible size of the effect.

The temperature in the laboratory compartment when the reactor was turned on was not observed to be changing within an accuracy of 1 - 2 degrees. It should be noted that the detector, together with a part of the electronic equipment, is inside a passive shield weighing 60 tons. This protection functions almost as a cryostat. The temperature coefficient of voltage stability at the PMT is 0.23 volts per degree. These studies were carried out in a special stand with a temperature change of 20 degrees. The change in the PMT Hamamatsu R5912 coefficient is 0.5% per degree, therefore, the spectrum shift in the 6000 keV region is 30 - 60 keV. Finally, in another special stand, the temperature stability of the entire system, including the PMT and the scintillator, had been checked. The temperature range was also 20 degrees, and the temperature stability was 1% per degree. Thus, the possible shift of the spectrum in the region of 6000 keV is 60 - 120 keV.

To test the possibility of generating a false oscillation effect, data processing was simulated taking into account such a bias. This processing showed that a shift of the spectrum in the region of 6000 keV by 120 keV is not dangerous and does not simulate the formation of a false oscillation effect.

4. The next question is an influence of unequal efficiency of neutrino events registration in different detector rows. It should be noted that in this case we are talking about the efficiency of correlated signals detection which involves several sections at once, as shown in FIG. 32 in section XIII on computer simulation of full-scale detector. Registration efficiency of the section depends on its position. However, each section moves together with its environment and its efficiency does not depend on the position of the detector. Here it is important to recall that we use the method of relative measurements with the movement of the detector and in addition, the registration efficiency for each section is canceled in $R_{ik}^{exp}$ ratio. If all sections collected data at every point, then no questions should arise at all. However, measurements by edge sections are not reproduced by central sections. As will be shown in the future, no such influence is observed beyond statistical accuracy. Nevertheless, a direct analysis of the possible impact of inequality among sections efficiencies on the observation of the oscillation process would be conducted.

The determination of the efficiency of individual sections should be done experimentally. In fact, the efficiency of the individual rows of the detector, which includes 5 sections located at the same distance, is needed.

For experimental determination of rows efficiency with respect to correlated signals, the background of fast neutrons of cosmic origin can be used. In principle, this requires a source that irradiates the entire detector uniformly, like a neutrino flux. The background of fast neutrons of cosmic origin is uniform as shown in FIG. 5 in section V. However, a self-screening process occurs in the detector, so the first and last rows of the detector show a higher count rate. The dependence of the count rate on the number of the detector row is shown in FIG. 48 at the top.

This dependency was obtained in the following measurements. The distances of detector movements correspond to section size (23.5 cm). All movements are controlled with laser distance measurer. The measurements were carried out at 10 detector positions in the way that the same distance from the reactor is measured with various detector rows. Spectra measured with various rows at same distance are averaged afterwards.

Average distribution of prompt signal counts obtained in background measurements during the whole period of reactor stop is shown in FIG. 48 (top). We should remind that first and last rows are not used for obtaining the final dependence on L/E. They are using as active and passive shielding from neutrons. They screen central fiducial part of the detector. It was mentioned before that direct rate (not correlated) of fast neutrons induced by cosmic background inside passive shielding is uniform. However, for the correlated background some gradient within the 6 – 12m distance range is observed, which can occur due to concrete structures of the building (see FIG. 2).

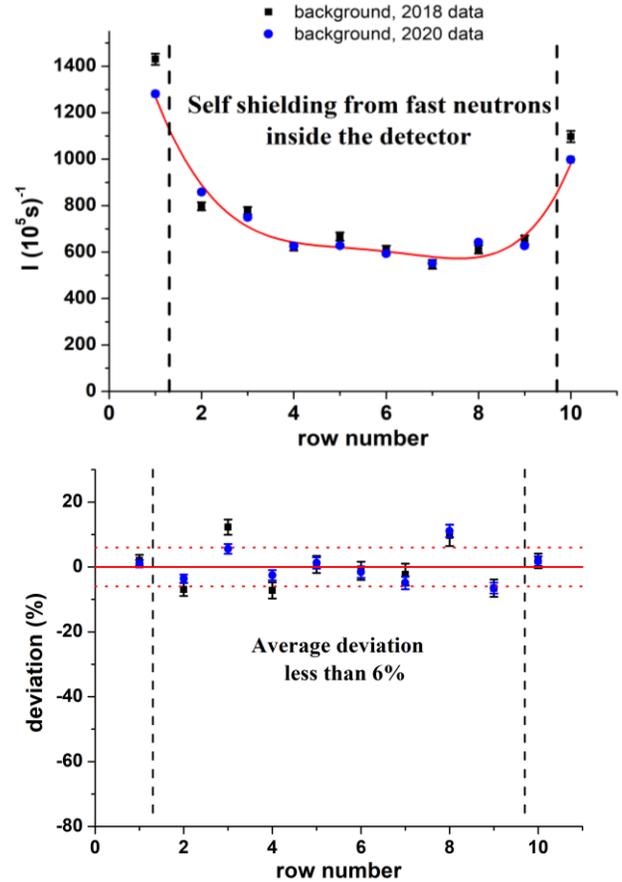

FIG. 48. Average distribution of correlated background prompt signals in detector over all positions (top). Deviation average distribution of prompt signals from profile. Profile was caused by inhomogeneity of fast neutrons background in the lab room (bottom).

Red line in FIG. 48 (top) is the profile of the fast neutron distribution due to self-screening effect. The deviation of counts from average value can be interpreted as difference in efficiency of different rows. The mean value of the deviation is ~ 6% (FIG. 48 bottom). However, different rows of the detector perform measurements at the same distance, and it is the reason why average efficiency that plays the role for particular distance

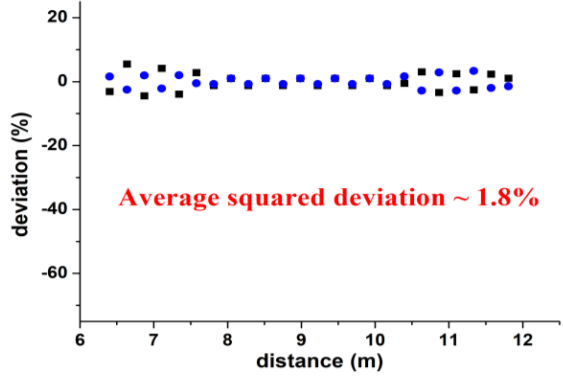
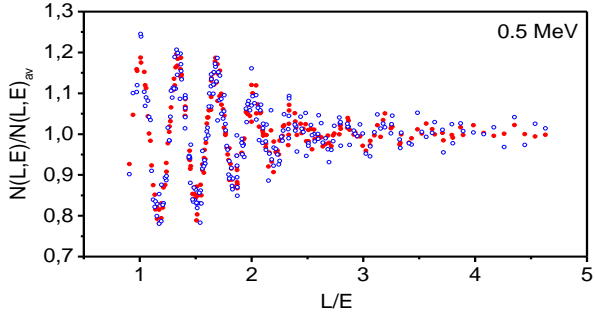
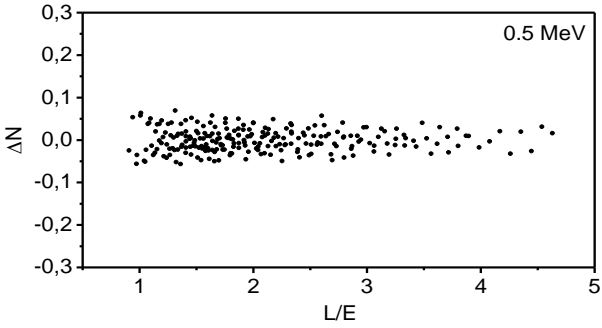
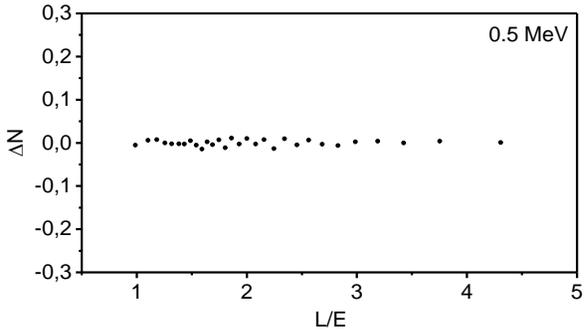

FIG. 49 Top – deviation of counts of correlated background of each distance from the reactor after averaging over rows from the mean value. Second – MC modeling of the oscillation effect considering deviations of the detection efficiencies for correlated events at different distances. Red dots – oscillation effect ignoring the influence of deviations of the correlated events registration efficiencies at different distances; blue dots – oscillation effect considering this influence. Third – difference between red and blue dots and fourth is sum by 8 of the third one.

To consider how differences in rows efficiencies affect the final results, one must take into account that averaging of spectra obtained with various rows at the same distance. In that approach the square deviation from the mean value is ~ 1.8%, as shown in FIG. 49 (top). It indicates that the detector inhomogeneity cannot be the origin of oscillation effect. Nevertheless, it is expedient to make MC simulations for complete clarity of the situation. Results of these simulations are on the FIG. 49 (second).

It can be seen, that considering deviations in the efficiency of detecting correlated events at different distances did not affect the effect of oscillations (see FIG. 49 third and fourth). This situation can be explained since the effect of oscillations is resonant and successfully survives in the presence of noise.

We can summarize our analysis of systematic uncertainties in conclusion that these uncertainties cannot explain the observed oscillation effect.

Indeed, after averaging over groups of eight points each, the effect of noise is significantly suppressed (see FIG. 49, fourth) because the characteristic noise frequency is higher than oscillation curve frequency. The amplitude of noise fluctuations is no more than 1% whereas the amplitude of the oscillation curve is 15%.

## XXIV. SYSTEMATIC ERRORS OF THE EXPERIMENT

One of possible systematic errors of oscillation parameter $\Delta m_{14}^2$ is determined by accuracy of energy calibration of the detector. The relative accuracy of ratio L/E is determined by the relative accuracy of measurements of energy, because the relative accuracy of measurements of distance is significantly better. The relative accuracy of measurements of energy in the most statistically significant area of the measured neutrino spectrum 3-4 MeV is 5%. Hence, possible systematic error of parameter $\Delta m_{14}^2$ is 0.6 eV², $\delta(\Delta m^2)_{syst1} \approx 0.6 eV^2$.

Another systematic error of parameter $\Delta m_{14}^2$ can occur in data analysis performed with $\chi^2$ method because of additional regions around the optimal value $\Delta m_{14}^2 \approx 7.3 eV^2$. In particular, the closest regions have values $5.2 eV^2$ and $8.8 eV^2$, as can be seen from the FIG. 43. However, contribution of the satellite at $5.2 eV^2$ is significantly smaller. Therefore, the possible systematic error of $\Delta m_{14}^2$ can be estimated as $\delta(\Delta m^2)_{syst2} \approx 1.0 \ eV^2$. Finally, the obtained value of difference squared masses is:

$\Delta m_{14}^2 = (7.3 \ \pm 0.13_{st} \pm 1.16_{syst})$ eV² $= (7.3 \pm 1.17)$ eV²

The statistical error of parameter $\sin^2 2\theta_{14}$ can occur in calculation of optimal value of $\sin^2 2\theta_{14}$ using $\chi^2$ method.

$$\sin^2 2\theta = 0.36 \pm 0.12_{stat}$$

## XXV. THE DEPENDENCE OF THE REACTOR ANTINEUTRINO FLUX ON DISTANCE IN RANGE 6-12 METERS

Results of measurements of the difference in counting rates of neutrino events (reactor ON-OFF) are shown in FIG. 50-52, as dependence of antineutrino flux on the distance to the reactor core. Fit of an experimental dependence with the law $A/L^2$ yields the following result. Goodness of that fit is 22%. However, there is an indication that in the range 9 – 12 m the rate is smaller than in the range 6 - 9 m and the curve that fitting the obtained points resemble the oscillation curve with period 6 – 8 m, which corresponds to $\Delta m_{14}^2 = 1.2$ eV$^2$. One can find the corresponding small green spot on the plane $\Delta m_{14}^2, \sin^2 2\theta_{14}$ (see FIG. 43 middle). However, this result is inconclusive because the deviation is within statistical uncertainties. Moreover, it can be the result of the fact that in the second measurement cycle the results were obtained mainly for the short distances. Finally, the background in range 9 – 12 m is partially suppressed by the concrete celling of the building. The statistical fluctuation is the most unbiased of all the mentioned hypotheses. Corrections for finite size of reactor core and detector sections are negligible – 0.3%, and correction for difference between detector movement axis and direction to center of reactor core is also negligible – about 0.6%.

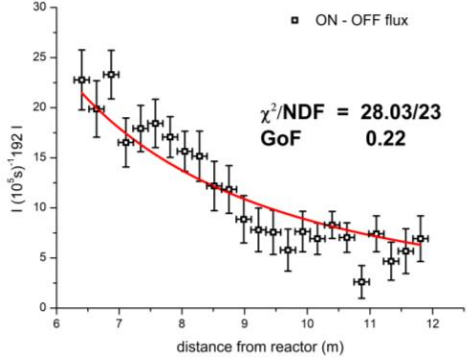

FIG. 50. Dependence of antineutrino flux on the distance to the reactor core – direct measurements with subtracted background

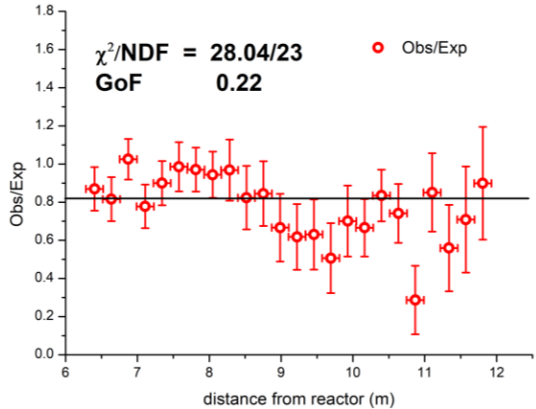

FIG. 51. Representation of experimental results in form of dependence of antineutrino flux on the distance to the reactor core normalized with the law $A/L^2$.

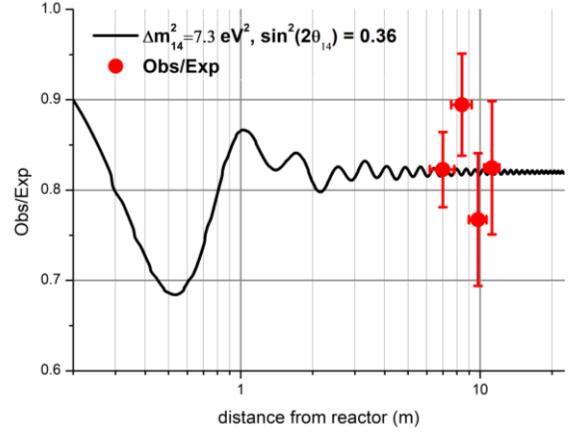

FIG. 52 Oscillation curve and experimental results in range 6-12 m.

## XXVI. MC MODEL OF THE EXPERIMENT WHICH TAKING INTO CONSIDERATION OBTAINED STATISTICAL ACCURACY

In section XVIII we presented the simulation of the experiment without taking into account the background conditions. In this section, we present the simulation of the experiment taking into account the background conditions observed in the experiment and the energy dependence of the energy resolution of the detector.

In the Monte Carlo model, as before, an antineutrino source with a reactor core size of 42x42x35 cm and an antineutrino detector consisting of 50 sections with geometric dimensions 22.5x22.5x75 cm were considered. Sections in one row have been combined. Antineutrinos were detected at all 24 distances from 6.5 m to 12 m with a row size step of 22.5 cm. The antineutrino spectrum was set in accordance with the antineutrino spectrum for $^{235}$U, although this is not important, since we used the method of relative measurements.

The number of simulated events corresponded to the experimental data (ON — OFF), accumulated in the experiment at different distances. So, the simulated count dispersion was in accordance with the statistical error of the experiment, so the background was taking into account. In this case we used statistical uncertainties obtained in the first and second cycle of measurements. Data collected during long time reactor OFF period were not included in the analysis. Thus, the experimental matrix $N_{ik} = N(E_i^\nu, L_k)$ was modeled. The ON — OFF neutrino count was simulated for two cases with and without oscillations. For the case with oscillations, we used the parameters obtained as a result of processing the experimental data — $\Delta m_{14}^2 = 7.3$ eV$^2$, $\sin^2 2\theta_{14} = 0.36$.

The final goal was to reproduce the experiment repeatedly and to form the distribution for the two mentioned cases by the data averaging method. To obtain an oscillatory curve, each of the simulated experiments was analyzed using the same processing scheme as in the real experiment. Each time the analysis was carried out with the hypothesis of oscillations and with the hypothesis without oscillations. Thus, it became clear which of the hypotheses is better suited for each of the experiments performed. For this, the value $\chi^2$ was calculated each time to fit the result with oscillation curve $\chi^2_{sin}$, or by the constant $\chi^2_{const}$. In general, $5 \cdot 10^6$ experiments were carried out for the case with oscillations and the same number without oscillations.

FIG. 53 shows the three-dimensional distribution of the reduced $\chi^2$ values on the plane $(\chi^2_{const}, \chi^2_{sin})$. It should be noted that the values $\chi^2_{const}$, $\chi^2_{sin}$ are correlated, since they were obtained from the same simulation, but with different processing hypothesis. It is quite obvious that for experiments without oscillations the optimal value of $\chi^2_{const} \approx 1$ should be obtained when processing with a hypothesis without oscillations. While when processing data with a hypothesis with oscillations, the optimal value will be $\chi^2_{sin} > 1$. For experiments with oscillations, the situation will be similar, but vice versa. Optimal for $\chi^2_{sin} \approx 1$, and $\chi^2_{const} > 1$. To generate these distributions 17 degrees of freedom and 19 degrees of freedom was used for analysis of simulations without oscillation and with oscillation, respectively. It should be noted that the distribution of results in the experiment was broadened (see section XII) and the simulation takes it into account.

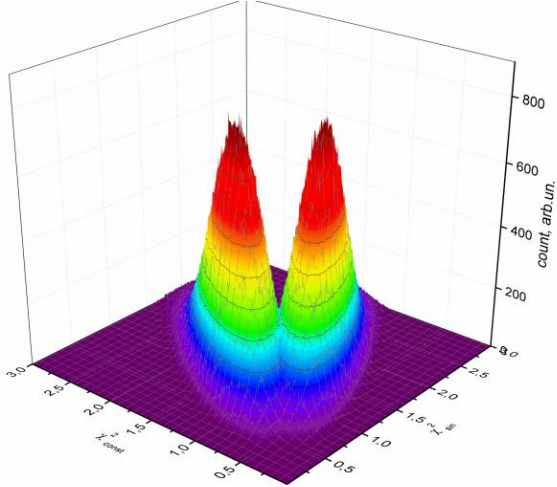 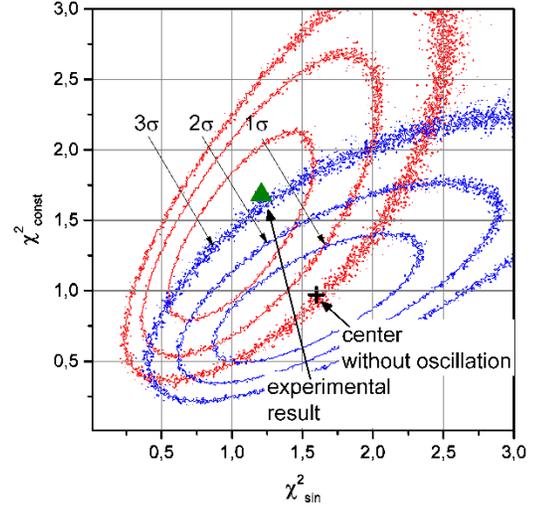

FIG. 53. On the left, the distribution of the reduced $\chi^2$ on the plane $(\chi^2_{const}, \chi^2_{sin})$ is shown for modeling with accuracy according to the experimental statistics and with a background level corresponding to the experimental background. Two cases were simulated with the hypothesis of oscillations and with the hypothesis without oscillations. For each case, an analysis was carried out with the hypothesis of oscillations and with the hypothesis without oscillations. On the right is the same picture, where the center of the distribution without oscillations (black cross) and the deviation from this center of the experimental result (green triangle – first and second cycle) are indicated. The result of the analysis of the experimental data when processed with the hypothesis with and without oscillations gives $\chi^2_{sin} = 1.21$ or 20.6 for 17 degrees of freedom and $\chi^2_{const} = 1.68$, or 31.9 for 19 degrees of freedom, respectively and $\Delta\chi^2 = 11.3$. Distribution contours with 1σ, 2σ and 3σ are marked. The deviation of the experimental result from the center of the distribution without oscillations is approximately 3σ.

The result of the simulation reveals that value $\Delta\chi^2 > 11.3$ corresponds to the probability of false signal of oscillation $2.6 \cdot 10^{-4}$, i.e. confidence level of oscillation effect with $\Delta m^2_{14} = 7.3$ eV$^2$, $\sin^2 2\theta_{14} = 0.36$ is 3.6 σ. However, the presented analysis of simulated data was done with fixed oscillation parameters. More rigorous analysis for possible accidental false oscillation effect could be done with free oscillation parameters.

As shown in [37,38], in order to obtain more accurate estimation of statistical significance one should use MC - simulation based statistical analysis. This method includes usage of free oscillation parameters and we applied it to analyze the results of the Neutrino-4 experiment. Within this approach we simulate spectra of prompt signals in the neutrino detector at different distances form the reactor using no oscillation hypothesis. Dispersion of simulated values was determined by statistical accuracy of $N_{ik}$ matrix of the experiment, i.e., by $\Delta N_{ik}$. Spectra of each simulation processed like experimental spectra, using the method described in section XVII, i.e. by constructing the matrix R. The simulation of values of the $N_{ik}$ matrix seems more accurate because such simulation better represents the order of data processing in our experiment. However, the calculations proves that the simpler approach based on direct usage of the matrix R gives similar final result. The values of $R_{ik}$ obtained from simulated values of $N_{ik}$ follows the normal distribution. We performed $1.5 \cdot 10^4$ simulations and the obtained distribution of $T = \chi^2_{null} - \chi^2_{best\,fit}$, where $\chi^2_{null} = \chi^2(0,0)$ – chi-squared for the null hypothesis, $\chi^2_{best\,fit} = \chi^2(\sin^2 2\theta, \Delta m^2)$ – minimal chi-squared corresponding to some oscillation parameters, is shown in fig. 54.

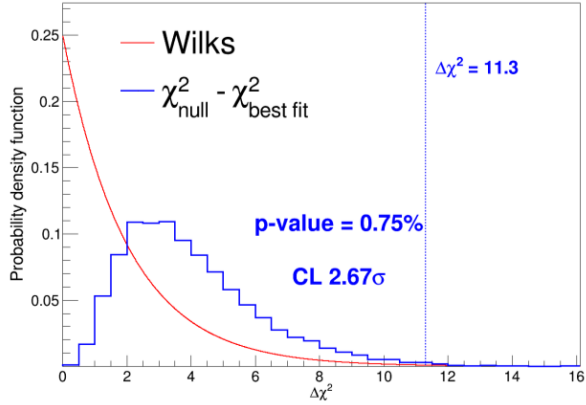

FIG. 54. $T$ distribution for MC based approach to the statistical analysis (blue line) and $\chi^2$ with 2 degrees of freedom function which is claimed by Wilks's theorem.

This distribution is different from $\chi^2$ with 2 degrees of freedom, which Wilks's theorem claims. However, the fraction of simulated results with $T \geq 11.3$ is only 0.0075, which corresponds to 2.7σ CL. Therefore, using this stricter criterion to estimate the confidence level in case of the oscillation search in the Neutrino-4 experiment we can make a conclusion about observation of neutrino oscillation into sterile state at 2.7σ CL.

We plan to continue data taking. However, it is quite obvious that the continuation of the collection of data without improving the installation is not advisable. It is necessary to suppress the background and improve the sensitivity of the setup. We are planning to improve the existing installation, and, in addition, we are planning to create a second neutrino laboratory at the SM-3 reactor with a detector that is 3 times more sensitive.

## XXVII. ANALISYS OF THE GENERAL SITUATION CONCERNING EXPERIMENTAL ATTEMPTS OF STERILE NEITRINO SEARCH

Despite the fact that the current confidence level is insufficient to make conclusion about existence of oscillation effect, it is reasonable to: 1) check if it contradict other results of experiments on search for sterile neutrino analyze the agreement with other experiments on search for sterile neutrino 2) analyze what experiments can confirm the obtained result. We do not aim to review other experiments. We only concern about the clear disagreement with other experiments which can impugn results of our experiment at current step. That is a very important part, because the issue in question is the discovery of a new particle outside the Standard Model. The detailed analysis was performed in the work [38]. Here we would like to present a short summary.

First of whole we should notice that this topic was investigated in many different experiments. In sector of the measurements at the reactors there are experiments performed at nuclear power plants and research reactors.

In experiments on nuclear power plants DANSS [16], NEOS [17] sensitivity to identification of effect of oscillations with large $\Delta m_{14}^2$ is considerably suppressed because of the big sizes of an active zone. The experiments PROSPECT [18] and STEREO [19] have good sensitivity to large values of $\Delta m_{14}^2$, but they have started data collection later, so possibly it will confirm or close our result in the future.

The oscillations obtained in the Neutrino-4 experiment gives oscillation parameter $\sin^2 2\theta_{14} \approx 0.36 \pm 0.12$ (2.9σ). The neutrino deficiency called gallium anomaly (GA) [7,8] gives oscillation parameter $\sin^2 2\theta_{14} \approx 0.32 \pm 0.10$ (3.2σ). The result of reactor antineutrino anomaly (RAA) [29-32] measurements is $\sin^2 2\theta_{14} \approx 0.13 \pm 0.05$ (2.6σ). Combination of these results gives an estimation for mixing angle $\sin^2 2\theta_{14} \approx 0.19 \pm 0.04$ (4.6σ).

The restrictions on the mixing angle which can be obtained from measurements of reactor anomaly and oscillations of the solar neutrinos are at level $\sin^2 2\theta_{14} \approx 0.1 - 0.2$ and contradict our result and the estimation obtained in measurements of gallium anomaly $\sin^2 2\theta_{14} \approx 0.3 - 0.4$. However, the discrepancy between Neutrino-4 and Reactor Antineutrino Anomaly is currently at 1.7σ level. Besides, the systematic uncertainties of the reactor and solar calculations remain questionable.

The values of oscillation parameters obtained in the Neutrino-4 experiment can be used to estimate the mass of the electron antineutrino, using general formulas for neutrino model [39, 40] with extension to 3+1 model.

Limitations on the sum of mass of active neutrinos $\sum m_\nu = m_1 + m_2 + m_3$ from cosmology are in the range $0.54 \div 0.11$ eV [41]. At the same time, knowing that $\Delta m_{14}^2 \approx 7.3$ eV$^2$, it is possible to consider that $m_4^2 \approx 7.3$ eV$^2$, and $m_1^2, m_2^2, m_3^2 \ll m_4^2$. Thus, the effective mass of the electron neutrino can be calculated by the formula: $m_{\nu_e}^{\text{eff}} \approx \sqrt{m_4^2 |U_{e4}|^2} \approx \frac{1}{2}\sqrt{m_4^2 \sin^2 2\theta_{14}}$.

It is necessary to make a little discussion here in connection with the known restrictions on the number of types of neutrinos and on the sum of the masses of active neutrinos from cosmology. Depending on the scale of masses, sterile neutrinos can influence the evolution of the Universe and be responsible for the baryonic asymmetry of the Universe and the phenomenon of dark matter [42]. However, the existence of the sterile neutrinos with low mass and mixing angle does not have a significant effect on cosmology [42]. Such sterile neutrinos practically do not thermalize in the primary plasma and leave it at an early stage.

Considering the facts mentioned above we can estimate the sterile neutrino mass $m_4 = (2.70 \pm 0.22)$ eV. In case of the parameter $\sin^2 2\theta_{14} \approx 0.19 \pm 0.04$ (4.6σ) obtained combining the results of the Neutrino-4 experiment and results of the Gallium anomaly measurements and more importantly using value $\Delta m_{14}^2 \approx (7.3 \pm 1.17)$ eV$^2$ obtained for the first time in the Neutrino-4 experiment, we can make an estimation of the electron neutrino mass: $m_{\nu_e}^{\text{eff}} = (0.59 \pm 0.11)$eV. Obtained neutrino mass does not contradict the restriction on neutrino mass $m_{\nu_e}^{\text{eff}} \leq 1.1$ eV (CL 90%) obtained in the KATRIN experiment [43].

In experiments for neutrinoless double beta decay, the Majorana neutrino mass is determined by the following expression: $m(0\nu\beta\beta) = \sum_{i=1}^{4}|U_{ei}|^2 m_i$ This expression for the model 3 + 1 and with $m_1, m_2, m_3 \ll m_4$ assumption can be simplified: $m(0\nu\beta\beta) \approx m_4 U_{14}^2$.. The averaged numerical value which includes the results of the Neutrino-4 and other experiments is: $m(0\nu\beta\beta) = (0.13 \pm 0.04)$eV. The best restrictions on the Majorana mass were obtained in the GERDA experiment [44]. Upper limit is $m_{\beta\beta} < [0.80 - 0.182]$eV. Therefore, our result does not contradict the results of the experiments on search for neutrinoless double beta decay.

The IceCube experiment and accelerator experiments LSND [1] and MiniBooNE [2] have the results that not only allow the sterile neutrino with parameter near $\Delta m_{14}^2 \approx 7.3$eV$^2$ to exist, but even can be considered as an indication on that possibility as shown in ref.[38].

## XXVIII. CONCLUSION

That work presents the steps of creating the experiment Neutrino-4 from the first preliminary experiments to the obtained result. The first results were published in works [23-28]. Here we discuss the data collected over 5 years to the present day. The first declaration of the observing the oscillation effect at the 3.5σ level was published in work [28]. Further measurements confirmed this result with statistical CL 3.2σ but taking into account possible systematics we reduce CL down to 2.9σ. More accurate statistical analysis gave stricter confidence level at 2.7σ. In June 2019 the SM-3 reactor was stopped for renovation. However, the background has been measured from July 2019 till October 2020. We tried to use additional background measurements, but it leads to increasing of statistical error due to uncompensated background fluctuations during long-term measurements.

As a result of the data analysis described in this work, we were able to estimate the actual energy resolution of the detector. Using the obtained energy resolution, we repeated the analysis of measurement results for first and second measurement cycles. The results are presented below:

$$\Delta m_{14}^2 = 7.30 \pm 0.13_{st} \pm 1.16_{syst} = 7.30 \pm 1.17.$$
$$\sin^2 2\theta = 0.36 \pm 0.12_{stat}$$

We plan to improve the current experimental facility of the Neutrino-4 experiment and also create the second neutrino laboratory at the SM-3 reactor which will be equipped with the three times more sensitive detector [45]. As a result, we will be able to obtain a final answer for the question of existence of the sterile neutrino with parameter $\Delta m_{14}^2$ near the 7eV$^2$.

## XXIX. ACKNOWLEDGEMENTS

The authors are grateful to the Russian Science Foundation for support under Contract No. 20-12-00079. Authors are grateful to M.V. Danilov, V.B.Brudanin, V.G.Egorov, Y.Kamyshkov, V.A.Shegelsky, V.V. Sinev, D.S. Gorbunov and especially to Y.G.Kudenko for beneficial discussion. Also, authors would like to thank C. Rubbia for useful questions. The delivery of the scintillator from the laboratory headed by Prof. Jun Cao (Institute of High Energy Physics, Beijing, China) has made a considerable contribution to this research.